\documentclass{elsart}

\usepackage{harvard}
\citationstyle{dcu}
\usepackage{graphicx}

\usepackage{amssymb}
\def\la{\mathrel{\hbox{\rlap{\hbox{\lower4pt\hbox{$\sim$}}}\hbox{$<$}}}}
\def\ga{\mathrel{\hbox{\rlap{\hbox{\lower4pt\hbox{$\sim$}}}\hbox{$>$}}}}

\newcommand{\bl}[1]{\mbox{\boldmath$ #1 $}}

\begin{document}

\begin{frontmatter}



\title{The applicability of the viscous $\alpha$-parameterization of gravitational 
instability in circumstellar disks}


\author[UWO,IF]{E. I. Vorobyov},
\ead{vorobyov@ap.smu.ca}

\address[UWO]{The Institute for Computational Astrophysics, Saint Mary's University, Halifax,
NS, B3H 3C3, Canada}
\address[IF]{The Institute of Physics, South Federal University, Stachki 194, Rostov-on-Don, Russia}

\begin{abstract}
We study numerically the applicability of the effective-viscosity approach for simulating
the effect of gravitational instability (GI) in disks of young stellar objects with different 
disk-to-star mass ratios $\xi$. We adopt two $\alpha$-parameterizations for the effective viscosity
based on \citeasnoun{Lin90} and \citeasnoun{Kratter} and compare the resultant disk
structure, disk and stellar masses, and mass accretion rates with those obtained
directly from numerical simulations of self-gravitating disks around low-mass 
($M_\ast\sim 1.0~M_\odot$) protostars. We find that the effective
viscosity can, in principle, simulate the effect of GI in stellar systems with 
$\xi \la 0.2-0.3$, thus corroborating a similar conclusion by \citeasnoun{Lodato04}
that was based on a different $\alpha$-parameterization. 
In particular, the Kratter et al's $\alpha$-parameterization has 
proven superior to that of Lin \& Pringle's,
because the success of the latter depends crucially on the proper choice of the 
$\alpha$-parameter.
However, the $\alpha$-parameterization generally fails in stellar systems
with $\xi \ga 0.3$, particularly in the Class 0 and Class I phases of stellar evolution, 
yielding too small stellar masses and too large disk-to-star mass ratios. In addition, 
the time-averaged mass accretion rates onto the star are underestimated
in the early disk evolution and greatly overestimated in the late evolution.
The failure of the $\alpha$-parameterization in the case of large $\xi$
is caused by a growing strength of low-order spiral modes in massive disks.
Only in the late Class II phase, when the magnitude of spiral modes diminishes and
the mode-to-mode interaction ensues, may the effective viscosity 
be used to simulate the effect of GI in stellar systems with $\xi\ga 0.3$.
A simple modification of the effective viscosity that takes into account 
disk fragmentation can somewhat improve the performance 
of $\alpha$-models in the case of large $\xi$ and even approximately reproduce the 
mass accretion burst phenomenon, the latter being a signature of the early gravitationally
unstable stage of stellar evolution \cite{VB2}.
However, further numerical experiments are needed to explore this issue.
\end{abstract}


\end{frontmatter}

\section{Introduction}
It has now become evident that circumstellar disks are prone to the 
development of gravitational instability in the early stage of 
stellar evolution \citeaffixed{Laughlin94,VB1,VB2,Krumholz,Kratter,Stamatellos08}{e.g.}. 
The non-axisymmetric spiral structure resulting from GI produces 
gravitational torques, which are negative in the inner disk and positive in the outer disk
and help limit disk masses via radial transport of mass and angular momentum \cite{Laughlin97,Adams89,VB1}.
Even in the late evolution, weak gravitational torques associated with 
low-amplitude density perturbations in the disk can drive
mass accretion rates typical for T Tauri stars \cite{VB5}.

The fact that gravitational torques trigger mass and angular momentum 
redistribution in the disk makes them conceptually similar to viscous torques, 
which are believed to operate in a variety of astrophysical disks. An 
anticipated question is then whether the GI-induced transport in
circumstellar disks can be imitated by some means of effective viscosity?
The answer depends on whether mass and angular momentum transport induced by 
gravitational torques is of global or local nature and this issue still remains 
an open question.

Lin \& Pringle \citeyear{Lin87,Lin90} were among the first to suggest that the transport induced 
by GI could be described within a viscous framework and parameterized the effect 
of GI using the following formulation for the effective
kinematic viscosity
\begin{equation}
\nu_{\rm eff} = \left\{ \begin{array}{ll} \alpha_{\rm LP} 
\left( {Q^2_{\rm c} \over Q^2_{\rm T}}  -1\right) \left( {c^2_{\rm s} \over \Omega} \right)
    & \,\, \mbox{if $Q_{\rm T} \le Q_{\rm c}$} \\
    0 
    & \,\, \mbox{otherwise},   \end{array}
   \right. 
\label{LinPringle}   
\end{equation}
where $\Omega$ is the angular speed, $c_{\rm s}$ is the sound speed, 
$Q_{\rm T}=c_{\rm s} \Omega/(\pi G \Sigma)$ is the Toomre parameter, $\Sigma$ is the gas 
surface density, and $Q_{\rm c}$
is the critical value of $Q_{\rm T}$ at which the disk becomes unstable against nonaxisymmetric 
instability. The dimensionless number $\alpha_{\rm LP}$ 
represents the efficiency of mass and angular momentum transport by GI.
It is evident that \possessivecite{Lin90} parameterization
is actually that of \possessivecite{Shakura}, with $\alpha$-parameter modified by the term 
$\left( Q^2_{\rm c}/Q^2_{\rm T} -1 \right)$ to describe the effect of GI. 

\citeasnoun{Laughlin96} have compared the evolution of a thin, self-gravitating 
protostellar disk using two-dimensional hydrodynamic simulations with the evolution
of a one-dimensional, axisymmetric viscous disk using the model of 
\citeasnoun{Lynden}. 
They suggest the following effective $\alpha$-parameter as a modification to the usual 
Shakura \& Sunyaev $\alpha$-model
\begin{equation}
\alpha_{\rm LR}= {\beta \Omega^\zeta \over \Sigma},
\label{laugh}
\end{equation}
where the surface density $\Sigma$ and angular velocity $\Omega$ are in specific
units defined in \citeasnoun{Laughlin96}, and $\zeta$ and $\beta$
are defined as
\begin{eqnarray}
\zeta &=& 0.325 - 2.25 \left( {M_\ast\over M_\ast+M_{\rm d}}\right), \\
\beta &=& 0.875 - 4.34 \gamma,
\end{eqnarray}
where $M_\ast$ and $M_{\rm d}$ are the star and disk masses, respectively,
and $\gamma$ is the polytropic constant. By considering models with 
different masses of protostellar disks and  different values of $Q_{\rm T}$, they concluded
that while their parameterization works better than that of Shakura \& Sunyaev, 
the $\alpha$-prescription fails in relatively massive disks with $M_{\rm d}\ga0.5~M_\ast$
due to the presence of global modes which are not tightly wound. 
\possessivecite{Laughlin96} parameterization also performs badly in low-mass disks
with $M_{\rm d} \la 0.2~M_\ast$, overestimating the actual radial mass transport 
due to gravitational torques.
In the intermediate mass regime, however, their parameterization can reproduce the
actual gas surface density profiles with a good accuracy.
The appearance of dimensional units in
the dimensionless coefficient $\alpha_{\rm LR}$ restricts the applicability of 
equation~(\ref{laugh}) -- one has to either use the 
\possessivecite{Laughlin96} system of units or redefine equation~(\ref{laugh}) according 
to the adopted units.

\citeasnoun{Lodato04} have studied the applicability of the viscous treatment 
of the circumstellar disk evolution.
They have modelled the evolution of self-gravitating circumstellar disks 
with masses ranging 
from $5\%$ to $25\%$ of the star using adiabatic equation of state with $\gamma = 5/3$ 
and cooling that removes energy on timescales of 7.5 orbital periods.
By calculating the gravitational and Reynolds stress 
tensors and assuming that the sum of these stresses is proportional to the gas pressure, 
they estimated the effective $\alpha$-parameter associated with gravitational and Reynolds stresses
to be $\alpha_{\rm g,R} \le 0.06$ in their numerical simulations.
By further assuming that their disks quickly settle in thermal equilibrium
(when the rate of viscous energy dissipation is balanced by radiative cooling), they
proposed the following expression for the saturated value of the 
$\alpha$-parameter \citeaffixed{Gammie}{see also}
\begin{equation}
\alpha_{\rm eq}= \left| {d \ln \Omega \over d \ln r} \right|^{-2} {1 \over \gamma(\gamma-1)t_{\rm cool}\Omega},
\end{equation}
where $t_{\rm cool}=7.5 \Omega^{-1}$ is a characteristic disk cooling time. 
For the parameters adopted in \possessivecite{Lodato04} numerical simulations, $\alpha_{\rm eq}=0.05$.
A good agreement between numerically derived
$\alpha_{\rm g,R}$ and $\alpha_{\rm eq}$ allowed them to conclude that the viscous treatment of self-gravity
is justified in disks with disk-to-star mass ratios $\xi\le0.25$ and, perhaps, even 
in more massive disks \cite{Lodato05}. 

Recently, \citeasnoun{Kratter} have suggested a modification to the usual 
Shakura \& Sunyaev $\alpha$-model based on the previous numerical simulations of \citeasnoun{Laughlin96},
Lodato \& Rice \citeyear{Lodato04,Lodato05}, and \citeasnoun{Gammie}.
Their $\alpha$-parameter invoked to simulate the effect of GI
($\alpha_{\rm GI}$) consists of two components: the ``short''
component $\alpha_{\rm short}$ and ``long'' component $\alpha_{\rm long}$
\begin{equation}
\alpha_{\rm GI}=\left(  \alpha^2_{\rm short} + \alpha^2_{\rm long} \right)^{1/2},
\label{KMK}
\end{equation}
where
\begin{eqnarray}
\alpha_{\rm short}&=&{\rm max}\left[0.14 \left({1.3^2\over Q_{\rm T}^2} -1 \right) \left(1-\mu\right)^{1.15},0  \right] 
 \label{Kratter1} \\
\alpha_{\rm long}&=&{\rm max}\left[ {1.4\times 10^{-3} (2-Q_{\rm T})\over \mu^{5/4} Q_{\rm T}^{1/2}
},0 \right],
\label{Kratter2}
\end{eqnarray}
where $\mu$ is the disk-to-total mass ratio.
The ``short'' component $\alpha_{\rm short}$ differs from the effective $\alpha$-parameter suggested
by Lin \& Pringle's equation~(\ref{LinPringle}) only in a mild $\mu$ dependence. 
The ``short'' and ``long'' components are meant to represent the different 
wavelength regimes of the gravitational instability expected to dominate at different values 
of $\mu$ and $Q_{\rm T}$.
We note that Kratter et al. also assumes $Q_{\rm T}={\rm max}(Q_{\rm T},1)$.

The use of the $\alpha$-model as a proxy for GI-induced transport has been challenged
by \citeasnoun{Balbus99}, who argue that the energy flux of self-gravitating disks is not
reducible to a superposition of local quantities such as the radial drift velocity  and stress
tensor -- the essence of an $\alpha$-disk. Instead, extra terms of non-local nature are present 
that mostly invalidate the $\alpha$-approach.
Their conclusion is corroborated by recent numerical simulations of collapsing massive cloud cores
by \citeasnoun{Krumholz},
who find that gravitational instability in embedded circumstellar disks with masses of 
order $50\%$ that of the central star is dominated by the $m=1$ spiral mode induced by
the SLING instability \cite{Adams89,Shu90}. This mechanism is global and
enables mass and angular momentum transport on dynamical rather than viscous timescales.

Circumstellar disks form in different physical environments and go through
different phases of evolution so that it is quite likely that
there is no unique answer on whether GI-induced transport can be described by some means of 
effective viscosity. Indeed, the early embedded phases of disk evolution (Class~0/I) 
are substantially influenced by an infalling envelope, both through
the mass deposition \cite{VB2,Kratter} and envelope irradiation \cite{Matzner05,Cai08}.
Younger Class~0/I disks are usually more massive than in the older Class~II ones 
\cite{Vor09}. 
As a result, gravitational instabilities in Class 0/I disk may be 
dominated by low-order ($m\le 2$), large-amplitude global modes \citeaffixed{Krumholz}{see e.g.}, 
which are likely to invalidate the viscous approach. 
On the other hand, Class~II disks evolve in relative isolation and settle into a steady state 
that is characterized by low-amplitude, high-order modes $m\ge 3$ \citeaffixed{Lodato04,VB3}{e.g.}.
Such modes tend to produce more fluctuations and cancellation in the net gravitational 
torque on large scales, thus making the viscous approach feasible.

The mass and angular momentum transport in self-gravitating disks 
is difficult to deal with analytically due to a kaleidoscope of competing spiral modes. 
Furthermore, long-term multidimensional numerical simulations of the evolution 
of circumstellar disks involving an accurate calculation of disk self-gravity  
are usually  very computationally intensive.
On the other hand, the theory of viscous accretion disks is fairly well developed 
\citeaffixed{Pringle}{e.g.} and is relatively easy to deal with numerically. 
This motivated many authors to use the viscous approach 
to mimic the effect of gravitational instability when studying the long-term evolution
of circumstellar disks
\citeaffixed{Lin90,Nakamoto95,Hueso05,Dullemond06,Kratter}{e.g.}.
It is therefore important to know if and when the viscous approach is applicable.
Some work in this direction has already been done by Lodato \& Rice 
\citeyear{Lodato04,Lodato05} and \citeasnoun{Cossins}, who considered
a short-term evolution of isolated disks with different disk-to-star mass 
ratios meant to represent different stellar evolution phases. 
However, they have not considered a long-term disk evolution due to 
an enormous CPU time demand of fully three-dimensional simulations. 
In the present paper we explore the applicability of the $\alpha$-parameterization
of gravitational instability along the entire stellar evolution sequence,
starting from a deeply embedded Class~0 phase and ending with a late Class II phase 
(T~Tauri phase).
Although the T Tauri phase of stellar evolution is likely to harbour only 
marginally gravitationally unstable disks with associated torques of 
low intensity \cite{VB3}, the disk structure 
may bear the imprints of the early, GI-dominated phase. That is why it is important 
to capture the main stages of disk evolution altogether in one numerical simulation.
We focus on the $\alpha$-parameterizations of \citeasnoun{Lin90} and \citeasnoun{Kratter} 
and defer a study of the \possessivecite{Lodato04} parameterization 
to a follow-up paper. Contrary to other studies, we run our numerical simulations of circumstellar
disks first with self-gravity calculated accurately by solving the Poisson integral 
and then with self-gravity imitated by effective viscosity. We then 
perform a detailed analysis of the resultant circumstellar disk structure, masses,
and mass accretion rates in the both approaches.

\section{Description of numerical approach}
\subsection{Main equations}
We seek to capture the main evolution phases of a circumstellar disk altogether,
starting from the deeply embedded Class~0 phase and ending with the T Tauri phase.
This can be accomplished only by adopting the so-called thin-disk approximation.
In this approximation, the basic equations for mass and momentum transport are 
written as \cite{VB2,VB4}
\begin{equation}
\label{visc1}
 \frac{{\partial \Sigma }}{{\partial t}} =  - \nabla _p  \cdot \left( \Sigma \bl{v}_p 
\right), 
\end{equation}
\begin{equation}
\label{visc2}
 \Sigma \frac{d \bl{v}_p }{d t}   =   - \nabla _p {\mathcal P}  + \Sigma \, \bl{g}^{\rm d}_p +
  \Sigma \, \bl{g}^{\rm s}_p
 + \left( \nabla \cdot \mathbf{\Pi}\right)_p \, ,
\end{equation}
where $\Sigma$ is the mass surface density, ${\mathcal P}=\int^{Z}_{-Z} P dz$ is the vertically integrated
form of the gas pressure $P$, $Z$ is the radially and azimuthally varying vertical scale height,
 $\bl{v}_p=v_r \hat{\bl r}+ v_\phi \hat{\phi}$ is the velocity in the
disk plane, $\nabla_p=\hat{\bl r} \partial / \partial r + \hat{\phi} r^{-1} 
\partial / \partial \phi $ is the gradient along the planar coordinates of the disk,
and $\bl{g}^{\rm d,s}_p=g^{\rm d,s}_r \hat{\bl r} +g^{\rm d,s}_\phi \hat{\phi}$ is the gravitational acceleration in the disk plane. The latter consists (in general) of two parts: that due to the disk
self-gravity ($\bl{g}^{\rm d}_p$) and that due to the gravity of the central star ($\bl{g}^{\rm s}_p$).
The gravitational acceleration $\bl{g}^{\rm d}_p$ 
is found by solving for the Poisson integral \cite{VB2}. 
The viscous stress tensor $\mathbf{\Pi}$ is expressed as
\begin{equation}
\mathbf{\Pi}=2 \Sigma\, \nu_{\rm eff} \left( \nabla v - {1 \over 3} (\nabla \cdot v) \mathbf{e} \right),
\end{equation}
where $\nabla v$ is a symmetrized velocity gradient tensor, $\mathbf{e}$ is the unit tensor, and
$\nu_{\rm eff}$ is the effective kinematic viscosity. 
The components of $(\nabla \cdot \mathbf{\Pi})_p$ in polar coordinates ($r,\phi$) 
can be found in \citeasnoun{VB4}. 
We emphasize that we do not take any simplifying assumptions about 
the form of the viscous stress tensor, apart from those imposed by the adopted 
thin-disc approximation. It can be shown \cite{Lodato08} that equation~(\ref{visc2}) can
be reduced to the usual equation for the conservation of angular momentum of a radial annulus 
in the axisymmetric viscous accretion disc \cite{Pringle}.

Equations~(\ref{visc1}) and (\ref{visc2}) are closed with a barotropic equation
that makes a smooth transition from isothermal to adiabatic evolution at 
$\Sigma = \Sigma_{\rm cr} = 36.2$~g~cm$^{-2}$:
\begin{equation}
{\mathcal P}=c_s^2 \Sigma +c_s^2 \Sigma_{\rm cr} \left( \Sigma \over \Sigma_{\rm cr} \right)^{\gamma},
\label{barotropic}
\end{equation}
where $c_s=0.188$~km~s$^{-1}$ is the sound speed in the beginning of numerical simulations 
and $\gamma=1.4$. Assuming a local vertical hydrostatic equilibrium in the disk,
$\Sigma_{\rm cr}=36.2$~g~cm$^{-2}$ becomes equivalent to the critical number
density $10^{-11}$~cm$^{-3}$ \cite{Masunaga00}.

The thin-disk approximation is an excellent means to calculate the evolution
for many orbital periods and many model parameters. It is well justified as long as
the aspect ratio $A\equiv A(r)$ of the disk vertical scale height $Z$ to radius $r$ 
does not considerably 
exceed 0.1.  The aspect ratio $A(r)$ for a Keplerian disk is usually approximated 
by the following expression 
\begin{equation}
A \le C \, Q_{\rm T} M_{\rm d}(r)/M_\ast, 
\end{equation}
where $M_{\rm d}(r)$
is the disk mass contained {\it within} radius $r$ and $C$ is a constant, the actual value of
which depends on the gas surface density distribution $\Sigma$ in the disk. For a disk
of constant surface density, $C$ is equal unity. However, circumstellar disks are characterized by 
surface density profiles declining with radius. For the scaling $\Sigma \propto r^{-1.5}$ 
typical for our disks, $C=1/4$. Adopting further $Q_{\rm T}=2.0$ and $M_{\rm d}(r)/M_\ast=0.5$,
which are the upper limits in our numerical simulations, we obtain the maximum value of
$A(r)=0.25$. This analysis demonstrates that the thin-disk approximation may be only 
marginally valid in the outer regions of a circumstellar disk, but is certainly justified 
in its inner regions
where $M_{\rm d}(r)/M_\ast$ is small. A typical distribution of the aspect ratio $A$ in a disk 
around one solar mass star was shown in figure 7 in \citeasnoun{Vor09}.

\subsection{Initial conditions}
\label{initial}
We start our numerical integration in the pre-stellar phase, which is
characterized by a collapsing (flat) starless cloud core, and continue into
the late accretion phase, which is characterised by a protostar-disk
system. This ensures a self-consistent formation of a circumstellar
disk, which occupies the innermost portion of the 
computational grid, while the infalling envelope (in the embedded stage of stellar 
evolution) occupies the rest of the grid.

We consider model cloud cores with mass $M_{\rm cl}=1.5~M_\odot$, initial 
temperature $T=10~K$, 
mean molecular weight $2.33$, and the outer radius $r_{\rm out}=0.07$~pc. 
The initial radial surface density and angular velocity
profiles are characteristic of a collapsing axisymmetric magnetically 
supercritical core \cite{Basu}:
\begin{equation}
\Sigma={r_0 \Sigma_0 \over \sqrt{r^2+r_0^2}}\:,
\label{dens}
\end{equation}
\begin{equation}
\Omega=2\Omega_0 \left( {r_0\over r}\right)^2 \left[\sqrt{1+\left({r\over r_0}\right)^2
} -1\right],
\end{equation}
where $\Omega_0$ is the central angular velocity.
The scale length $r_0 = k c_s^2 /(G\Sigma_0)$, where $k= \sqrt{2}/\pi$
and $\Sigma_0=0.12$~g~cm$^{-2}$. 
These initial profiles are characterized by the important
dimensionless free parameter $\eta \equiv  \Omega_0^2r_0^2/c_s^2$
and have the property that the asymptotic ($r \gg r_0$) ratio of centrifugal to gravitational
acceleration has magnitude $\sqrt{2}\,\eta$ \cite{Basu}. 
The centrifugal radius of a mass shell initially located at radius $r$ is estimated to be
$r_{\rm cf} = j^2/(Gm) = \sqrt{2}\, \eta r$, where $j=\Omega r^2$ is the specific angular
 momentum.


The strength of gravitational instability is expected to depend on the disk mass. 
According to Lodato \& Rice (2004,2005), a viscous parameterization of GI is allowed
in systems with the disk-to-star mass ratio $\xi \la 0.25$ but may fail in systems with relatively more
massive disks. 
It is therefore important to consider systems with different disk-to-star mass ratios. This can 
be achieved by varying the initial rate of rotation of a model cloud core,
but keeping all other cloud core characteristics fixed. Indeed, an increase in $\Omega_0$
would result in a larger value of $\eta$ and $r_{\rm cf}$.
In turn, an increase in the centrifugal radius $r_{\rm cf}$ would result 
in more mass landing onto the disk rather than
being accreted directly onto the central star (plus some inner circumstellar region at $r<5$~AU
which is unresolved in our numerical simulations), thus raising the resultant disk-to-mass ratio.
According to \citeasnoun{Caselli},
velocity gradients in dense molecular cloud cores range between 0.5 and 6.0 km~s$^{-1}$~pc$^{-1}$.
Therefore, we choose three typical values for the central angular velocity of our model
cloud cores: $\Omega=\Omega_{0,1}=0.82$~km~s$^{-1}$~pc$^{-1}$, $\Omega=\Omega_{0,2}=1.5$~km~s$^{-1}$~pc$^{-1}$,
and $\Omega=\Omega_{0,3}=2.5$~km~s$^{-1}$~pc$^{-1}$. It is convenient to parameterize our models in terms of
the ratio of rotational to gravitational energy $\beta$, which is very similar 
in magnitude to the parameter $\eta$ introduced above. \citeasnoun{Caselli}
report $\beta$ ranging between $10^{-4}$ and $0.07$ in their sample of dense molecular cloud cores.
Our model cloud cores have $\beta_1=8.83\times 10^{-4}$, $\beta_2=2.3\times 10^{-3}$, 
and $\beta_3=8.2\times 10^{-3}$. For the sake of conciseness, we present the results of the 
$\beta_1$ and $\beta_3$ models, referring to the intermediate $\beta_2$ model only where necessary.

Equations~(\ref{visc1}), (\ref{visc2}), (\ref{barotropic}) are solved in polar 
coordinates $(r, \phi)$ on a numerical grid with
$128 \times 128$ cells. We use the method of finite differences with a time-explicit,
operator-split solution procedure. Advection is
performed using the second-order van Leer scheme.  The radial points are logarithmically spaced.
The innermost grid point is located at $r_{\rm in}=5$~AU, and the size of the 
first adjacent cell is 0.3~AU.  We introduce a ``sink cell'' at $r<5$~AU, 
which represents the central star plus some circumstellar disk material, 
and impose a free inflow inner boundary condition. The sink cell is dynamically inactive but 
serves as the source of gravity, thereby influencing the gas dynamics in the active
computational grid. The outer boundary is reflecting.
The gravity of a thin disk is computed by directly summing the input from each computational cell
to the total gravitational potential. The convolution theorem is used to speed up 
the summation. A small amount of artificial viscosity is added to the code to smear out
shocks over one computational zone according to the usual 
\possessivecite{vonNeumann} prescription, though the associated artificial viscosity torques 
were shown to be negligible in comparison with gravitational torques \cite{VB3}. 
A more detailed explanation of numerical methods and relevant tests 
can be found in Vorobyov \& Basu \citeyear{VB2,VB3}.

\subsection{Three numerical models}

We consider three numerical models that are distinct in the way the right-hand side of 
equation~(\ref{visc2}) is handled. In the first model (hereafter, self-gravitating model or SG model),
viscosity is set to zero throughout the entire evolution (forth term) and the system evolves
exclusively via gravity of the disk and central star (second and third terms, respectively), 
as well as pressure forces (first term). 
In the second model (hereafter, Lin and Pringle model or LP model), the disk self-gravity 
is switched off after the disk formation (no second term) and the subsequent
evolution is governed by pressure forces (first term), gravity of the central star (third term), 
viscosity (forth term). 
In particular, the kinematic viscosity is computed using the following representation
\begin{equation}
\nu_{\rm eff} = \left\{ \begin{array}{ll} {\alpha_{\rm LP}} 
\left( {Q^2_{\rm c} \over Q^2_{\rm T}}  -1\right)  {c_{\rm s} h} 
    & \,\, \mbox{if $Q_{\rm T} \le Q_{\rm c}$} \\
    0 
    & \,\, \mbox{otherwise},   \end{array}
   \right. 
   \label{LP}
\end{equation}
where $h$ is the disk scale height. The latter is calculated assuming the vertical 
hydrostatic equilibrium in the gravitational field of the disk and the central star 
\citeaffixed{Vor09}{see}. 
For numerical reasons, $\nu_{\rm eff}$ is set to zero if $Q_{\rm T}$
drops accidentally below some low value, which is set in our simulations to 0.3.
The third model (hereafter, KMK model) differs from the second model only in the way 
the effective viscosity is 
defined. More specifically, we use the following expression for the 
effective kinematic viscosity
\begin{equation}
\nu_{\rm eff}=\alpha_{\rm GI} c_{\rm s} h,
\end{equation}
where $\alpha_{\rm GI}$ is determined from equation~(\ref{KMK}).
Following \citeasnoun{Kratter}, we also set $Q_{\rm T}={\rm max}(Q_{\rm T},1)$.
In Section~\ref{fragment}, we consider a modification to the
KMK model that attempt to deal with the fragmentation regime at $Q_{\rm T}<1.0$. 
All the three numerical models start from identical initial conditions as described 
in Section~\ref{initial}

\section{Cloud cores with low rates of rotation.}
In this section we consider model cloud cores that are described by the ratio of rotational 
to gravitational energy $\beta\equiv\beta_1=0.88\times 10^{-3}$. 
This value is chosen to represent dense cloud cores with low rates of rotation, 
as inferred from the measurements of \citeasnoun{Caselli}.

\subsection{The LP model}
\label{LPlow}
\begin{figure}
  \resizebox{\hsize}{!}{\includegraphics{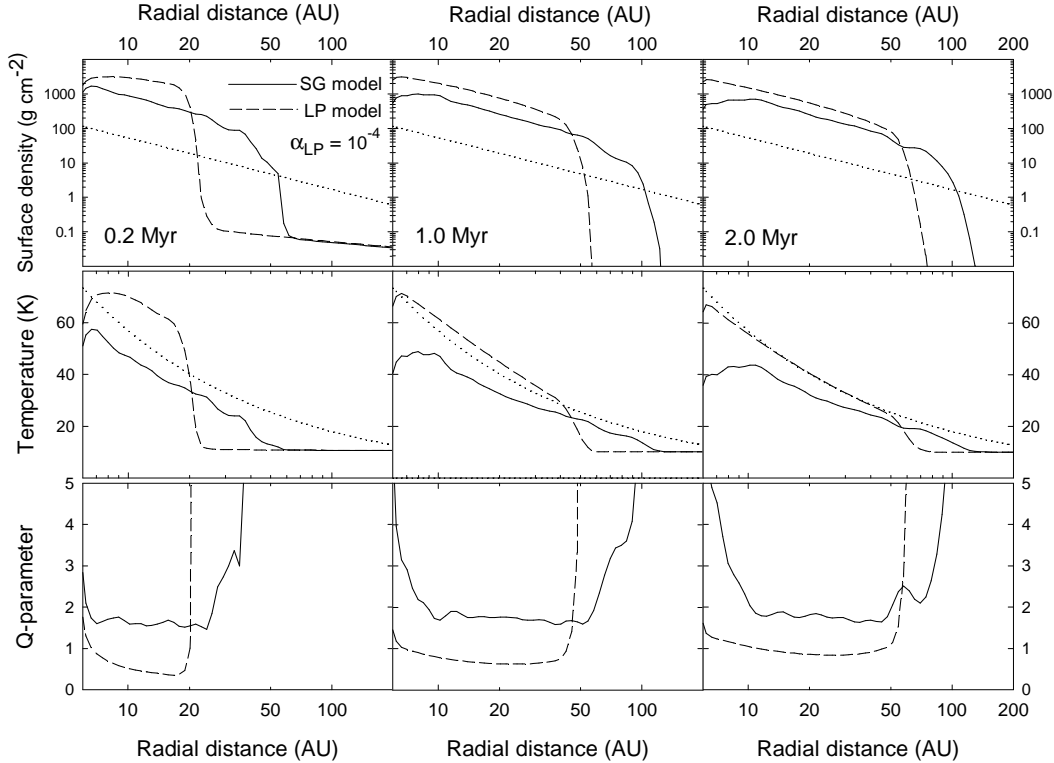}}
      \caption{Gas surface density (top row), gas temperature (middle row), and $Q$-parameter (bottom
      row) as a function of radius in the $\alpha_{\rm LP}=10^{-4}$ model (dashed lines) and 
      SG model (solid lines). Three
      distinct evolutionary times since the formation of the central protostar are indicated in each
      column. The dotted lines show the MMSN gas surface density profile (top row) and 
      a gas temperature profile typical for circumstellar disk (middle row) 
      as inferred by AW05.
       }
         \label{fig1}
\end{figure}

\begin{figure}
  \resizebox{\hsize}{!}{\includegraphics{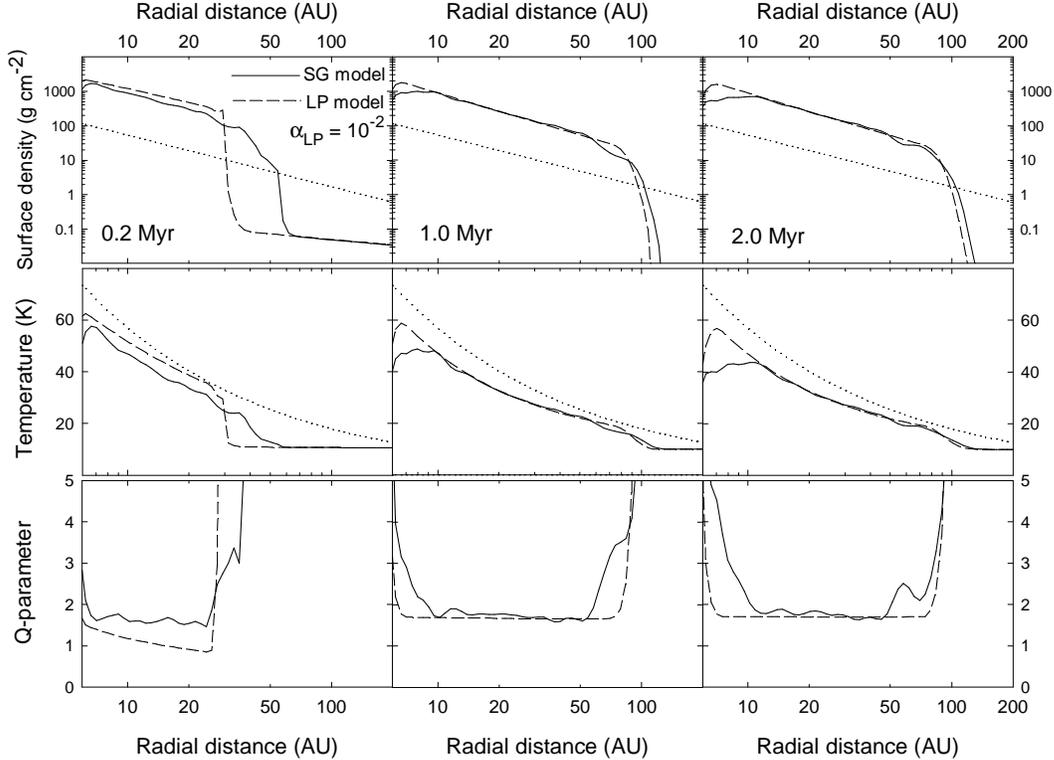}}
      \caption{The same as Fig.~\ref{fig1} only for the $\alpha_{\rm LP}=10^{-2}$~model.}
         \label{fig2}
\end{figure}

\begin{figure}
  \resizebox{\hsize}{!}{\includegraphics{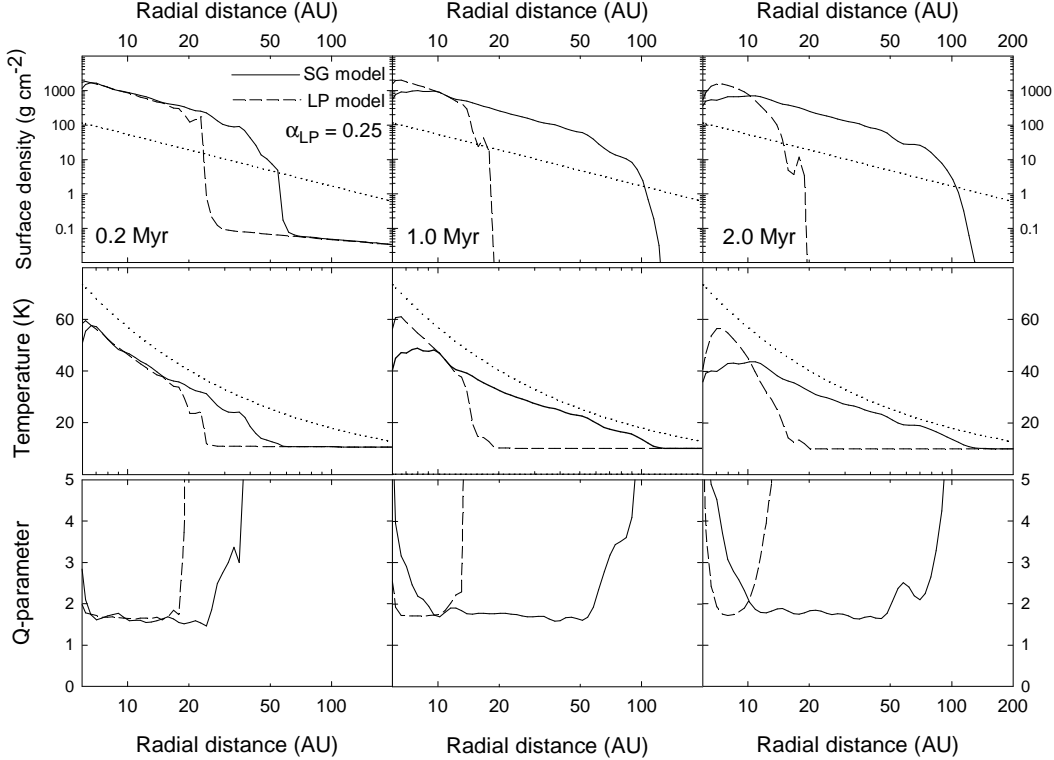}}
      \caption{The same as Fig.~\ref{fig1} only for the $\alpha_{\rm LP}=0.25$~model.}
         \label{fig3}
\end{figure}

Equation~(\ref{LP}) indicates that the LP model has two free parameters: $\alpha_{\rm LP}$
and $Q_{\rm cr}$. It is therefore our main purpose to determine the values of $\alpha_{\rm LP}$, 
using which the LP model reproduces best the exact solution provided by the 
SG~model. We have found that the LP~model depends only weakly on $Q_{\rm cr}$,
as soon as its value is kept near 1.5. In this study we set $Q_{\rm cr}=1.7$.
We vary $\alpha_{\rm LP}$ in wide limits, starting from $10^{-4}$ and ending with $0.5$. 
Figures~\ref{fig1}-\ref{fig3} show the disk radial structure obtained in the SG model 
(solid lines) and LP model (dashed lines). More specifically, the top panels show the radial
gas surface density distributions, while the middle and bottom panels show the radial profiles of
the gas temperature and Toomre parameter $Q_{\rm T}$, respectively. It may not be entirely 
consistent to calculate $Q_{\rm T}$ in the non-self-gravitating model. Nevertheless, this
quantity may serve as an indicator of how far the stability properties of the LP model 
deviate from the exact solution.
Three characteristic times based on the age of the central star in the SG~model
have been chosen: $t=0.2$~Myr (left row), $t=1.0$~Myr (middle row), and
$t=2.0$~Myr (right row). The left row represents an evolution stage when about 
$25\%$ of the initial cloud core material is still contained in the infalling 
envelope, while the middle and right rows represent a stage when the
envelope has almost vanishes. Three values of the $\alpha$-parameter 
have been selected for the presentation: $\alpha_{\rm LP}=10^{-4}$ (Fig.~\ref{fig1}), 
$\alpha_{\rm LP}=0.01$ (Fig.~\ref{fig2}), and $\alpha_{\rm LP}=0.25$ (Fig.~\ref{fig3}).
The dotted lines show the minimum mass solar nebular (MMSN)
density profile $\Sigma=1.7\times 10^{3} (r/1\mathrm{AU})^{-1.5}$ 
\cite{Hayashi} and a disk radial temperature
profile $T=180 (r/1\mathrm{AU})^{-0.5}$ inferred by \citeasnoun{Andrews05}
(hereafter AW05) from a large sample of YSO in the Taurus-Aurigae star-forming region.

It is instructive to review the main disk properties 
obtained by the SG~model. 
An accurate determination of disk masses in numerical simulations of collapsing 
cloud cores is not a trivial task. Self-consistently formed circumstellar disks 
have a wide range of masses and sizes, which are not known {\it a priori}. 
In addition, they often experience radial pulsations in the early
evolution phase, which makes it difficult to use such tracers as rotational support
against gravity of the central star. However, numerical and observational 
studies of circumstellar disks indicate that the disk
surface density is a declining function of radius. Therefore, we distinguish
between disks and infalling envelopes using a critical surface density for the 
disk-to-envelope transition, for which we choose a fiducial value of $\Sigma_{\rm tr}=0.1$~g~cm$^{-2}$.
This choice is dictated by the fact that densest {\it starless} cores have surface
densities only slightly lower than the adopted value of $\Sigma_{\rm tr}$.
In addition, our numerical simulations indicate that self-gravitating disks have 
sharp outer edges and the gas densities of order $0.01-0.1$~g~cm$^{-2}$ 
characterize a typical transition region between the disk and envelope.

As the solid lines in Figs~\ref{fig1}-\ref{fig3} indicate,
most of the self-gravitating disk is characterized by a power-law surface density 
distribution declining with radius as $r^{-1.5}$. This slope is also predicted 
by the MMSN hypothesis \cite{Hayashi}. 
However, our obtained gas surface densities are almost 
a factor of 10 greater than those of the MMSN throughout the entire 
disk evolution. This feature is an important property of self-gravitating disks 
\citeaffixed{VB4}{see also}. It makes easier giant planets to form, because 
planet formation 
scenarios seem to require gas densities at least a few times larger than those 
of the minimum-mass disk \cite{Ida04,Boss01}. The radial gas temperature profiles
indicate that the self-gravitating disk is somewhat colder than 
the typical disk in Taurus-Aurigae region (AW05). This is particularly true
in the late evolution, though large deviations from the typical profile toward 
colder disk are also present in the AW05 sample. We point out that irradiation by
a central source can raise the disk temperature and seriously affect the disk
propensity to fragmentation in the inner disk regions 
\citeaffixed{Matzner05}{e.g.}. Unfortunately, this effect is difficult
to take into account self-consistently in polytropic disks due to the lack
of detailed treatment of cooling and heating. Hotter disk can be obtained in our numerical 
simulations by choosing a larger value for the ratio of specific heats $\gamma$ 
in the barotropic equation of state (\ref{barotropic}). We discuss 
hotter disks in Section~\ref{hotdisk}. We also note that our self-gravitating 
disk exhibits a near-Keplerian rotation.

Finally, the time behaviour of the Toomre parameter $Q_{\rm T}$ in a self-gravitating 
disk warrants some attention. It is evident that 
the self-gravitating disk regulates itself near the boundary of gravitational stability,
with values of $Q_{\rm T}$ lying in the $1.7-2.0$ range.
This important property breaks down when a sufficient amount of {\rm physical} viscosity 
(i.e. due to turbulence) is present in the disk \citeaffixed{VB4}{see e.g.}.
We note that the $Q$-parameter in the early evolution of a self-gravitating disk 
may occasionally drop below $1.7-2.0$ in some parts of the disk. These episodes are 
usually associated with disk fragmentation and formation of dense clumps, 
which have masses of up to 10--20 Jupiters, sizes
of several AU, number densities of up to $10^{13}-10^{14}$~cm$^{-3}$ 
and are pressure supported against their own gravity.
Most of these clumps are quickly driven onto the central protostar
by gravitational torques from spiral arms. This phenomenon causes a burst of 
mass accretion \cite{VB1,VB2} and is very transient in nature (the burst itself
takes less than 100~yr). A small number 
of the clumps may be flung into the outer regions where they disperse,
most likely due to a combined action of tidal forces, differential rotation, and 
insufficient numerical resolution (our grid is logarithmically spaced in the radial direction).

We now proceed with comparing the disk structure in the SG and LP~models. 
Figures~\ref{fig1}-\ref{fig3} reveal that both the $\alpha_{\rm LP}=10^{-4}$ model
and $\alpha_{\rm LP}=0.25$ model fail to accurately reproduce the radial structure 
of the disk obtained in the SG model. More specifically, the $\alpha_{\rm LP}=10^{-4}$ model
yields too dense and hot disks throughout the entire evolution. 
The disagreement is particularly strong in the early disk evolution ($t=0.2$~Myr), when
the radial gas surface density distribution
becomes substantially shallower than that predicted by the exact solution 
($\propto r^{-1.5}$). 
The obtained disk radii in the $\alpha_{\rm LP}=10^{-4}$ model 
are smaller by a factor of $2-3$ than those found in the SG~model. 
The Toomre parameter is also considerably smaller than that of the self-gravitating disk.
In summary, the LP model with $\alpha_{\rm LP}$ as small as  $10^{-4}$ fails 
to provide an acceptable fit to the exact solution.

When we turn to large values of the $\alpha$-parameter $\alpha_{\rm LP}=0.25$ (Fig~\ref{fig3}), 
the corresponding LP~model 
seems to yield density and temperature profiles that are in acceptable agreement with those 
of the exact solution only in the very early phase of disk evolution ($t=0.2$~Myr). Even in
this early stage, the disk size is severely underestimated. Furthermore, the late evolution
sees a strong mismatch between the disk structure in the LP and SG~models. We conclude that
large values of $\alpha_{\rm LP} \ga 0.25$ may be marginally acceptable in the early, 
strongly gravitationally unstable phase of disk evolution,
but cannot be used to simulate the effect of self-gravity on long time scales of order of
several Myr.

The $\alpha_{\rm LP}=10^{-2}$ model (dashed lines in Fig.~\ref{fig2}) appears to provide 
the best fit 
to the exact solution. Although in the early disk evolution ($t=0.2$~Myr) 
the simulated gas surface densities and 
temperatures are somewhat larger than those provided by the SG~model, in the late 
evolution ($t\ga 1$~Myr) they are almost 
indistinguishable from the exact solution throughout most of the disk. 
We conclude that the LP model can be rather successful 
in reproducing the effect of
gravitational instability in star-disk systems formed form slowly rotating cloud cores, 
provided that the value of $\alpha_{\rm LP}$ lies close to $10^{-2}$.

\begin{figure}
  \resizebox{\hsize}{!}{\includegraphics{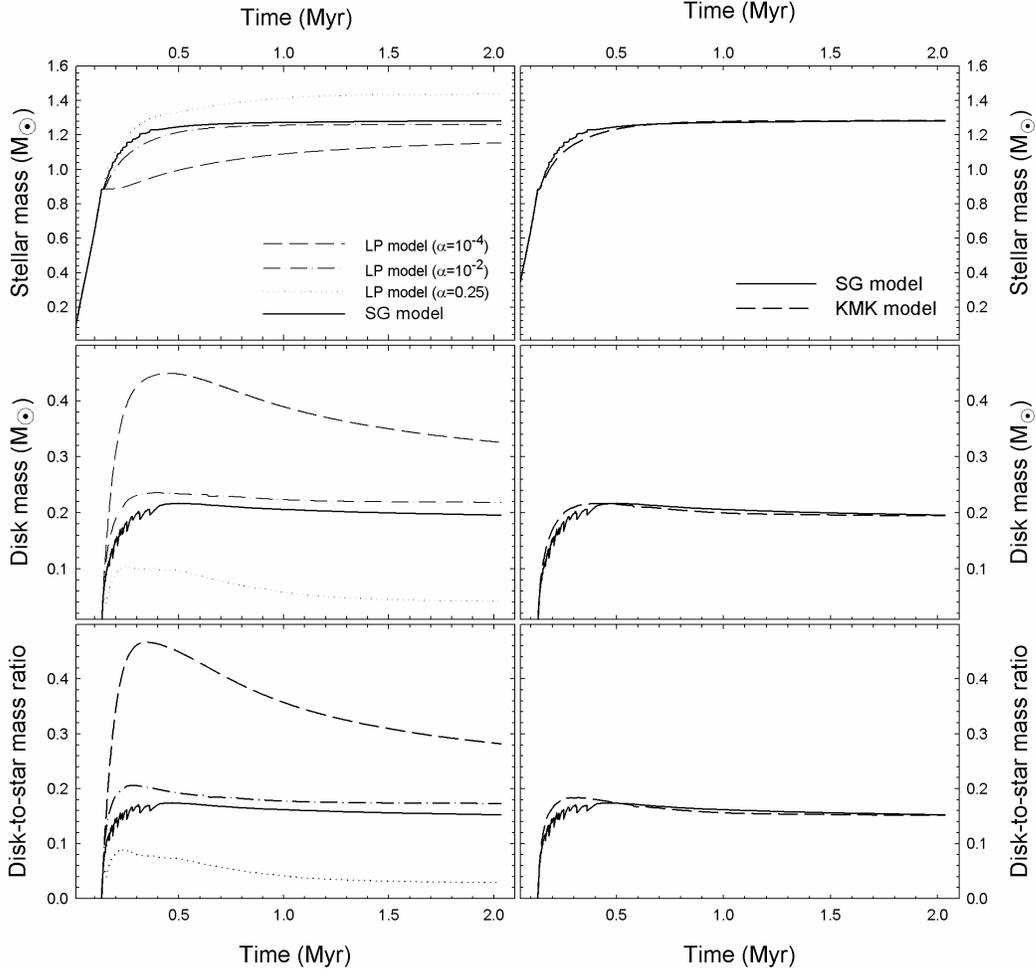}}
      \caption{Stellar masses (top row), disk masses (middle row), and disk-to-star mass ratios (bottom
      row) in the LP model (left column) and KMK model (right column). In particular, the dashed, dash-dotted,
      and dotted lines in the left column present data for the $\alpha_{\rm LP}=10^{-4}$ model,
      $\alpha_{\rm LP}=10^{-2}$ model, and $\alpha_{\rm LP}=0.25$ model, respectively,
      while the dashed line in the right column shows data for the KMK model. The solid lines in
      both columns correspond to the SG model.}
         \label{fig4}
\end{figure}

It is interesting to compare disk and stellar masses obtained in the LP models with
those of the SG model. The left column in Fig.~\ref{fig4} shows the stellar mass 
(top), disk mass (middle), and disk-to-star mass ratio $\xi$ (bottom) for the 
SG and LP models. More specifically, 
the solid, dashed, dash-dotted, and dotted lines present data for the SG~model, 
$\alpha_{\rm LP}=10^{-4}$ model, $\alpha_{\rm LP}=10^{-2}$ model, and
$\alpha_{\rm LP}=0.25$~model, respectively. We use $\Sigma_{\rm crit }=0.1$~g~cm$^{-2}$ for the 
disk-to-envelope transition. The horizontal axis shows time elapsed since the formation of the 
central star. The disk forms at $t=0.14$~Myr\footnote{In fact, the disk forms earlier but its evolution
is unresolved in the inner 5~AU due to the use of a sink cell in our numerical simulations. However,
the mass contained in this inner 5~AU is negligible compared to the rest of the resolved disk.}, 
when the central object has accreted approximately $60\%$ of the initial cloud core mass 
$M_{\rm cl}=1.5~M_\odot$. The disk-to-star mass ratio in the SG~model never exceeds $\xi=17\%$.
It is evident that the $\alpha_{\rm LP}=10^{-2}$ model yields the masses and disk-to-star 
mass ratios that agree best with those derived in the SG~model. 
The high-viscosity LP model ($\alpha_{\rm LP}=0.25$)  
predicts too large values for the stellar mass and too low values for the disk mass, 
whereas the low-viscosity LP model ($\alpha_{\rm LP}=10^{-4}$) does it vice versa.
This example nicely illustrates the sensitivity of the LP model to the choice of 
$\alpha_{\rm LP}$.

\begin{figure}
  \resizebox{\hsize}{!}{\includegraphics{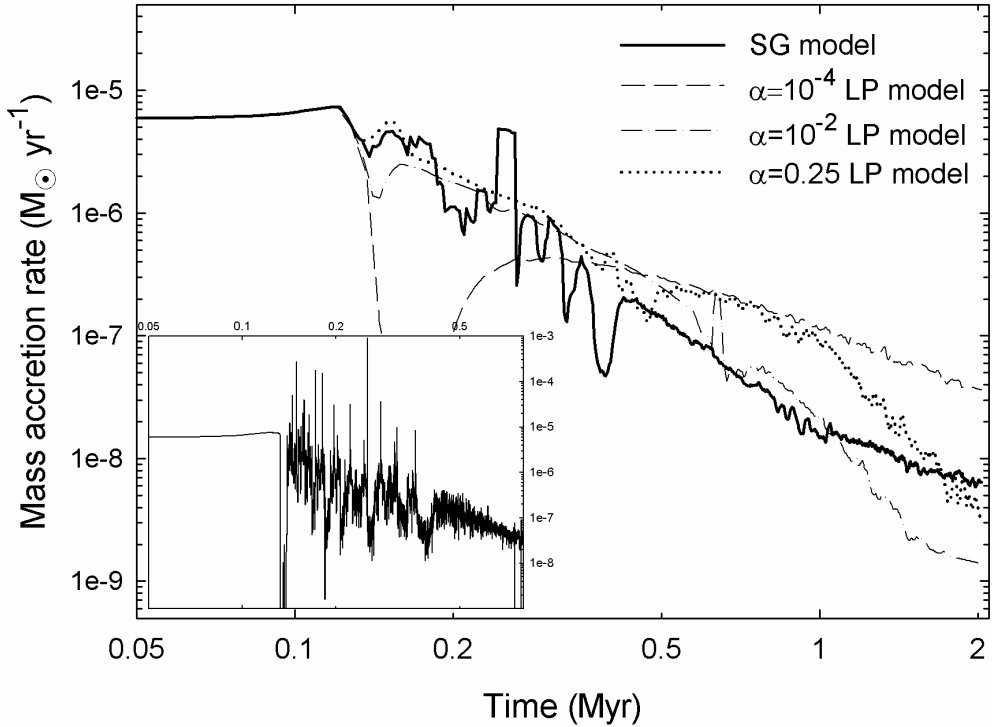}}
      \caption{Time-averaged mass accretion rates in the SG model (solid line), $\alpha_{\rm LP}=10^{-4}$
      LP model (dashed line), $\alpha_{\rm LP}=10^{-2}$ LP model (dash-dotted line), and 
      $\alpha_{\rm LP}=0.25$ LP model (dotted line). The insert shows the instantaneous 
      mass accretion rate versus time in the SG model.}
         \label{fig5}
\end{figure}

Finally, we calculate the instantaneous mass accretion rates through the inner 
disk boundary as $\dot{M}=-2\pi r_{\rm in} \Sigma \, v_{\rm r}$, where
$r_{\rm in}=5$~AU, and $v_{\rm r}$ is the radial gas velocity at $r_{\rm in}$. There is evidence 
that accretion on to the star 
is a highly variable process in the early embedded stage of disk evolution. 
Observations indicate that, in addition to young stellar objects (YSOs) with 
mass accretion rates similar to those predicted by \citeasnoun{Shu77}, 
there is a substantial populace of YSOs
with the sub-Shu accretion rates $\dot{M}\le 10^{-6}~M_\odot$~yr$^{-1}$ \cite{Enoch09}
and a small number of super-Shu accretors with $\dot{M}>10^{-5}~M_\odot$~yr$^{-1}$.
Furthermore, numerical simulations show that the disk in the early embedded phase 
becomes periodically destabilized due to the mass deposition from an 
infalling envelope {\cite{VB1,VB2,Boley09}} . The resulted gravitational torques drive 
excess mass in the form of dense clumps on to the central star, 
thus producing the so-called burst phenomenon \cite{VB1,VB2}. This phenomenon
can only be captured approximately by any model that mimics mass transport 
in self-gravitating disks via an effective viscosity \citeaffixed{Kratter}{see e.g.}.
In Section~\ref{fragment}, we try to reproduce the burst behaviour within the framework 
of the KMK model by modifying the definition of the $\alpha$-parameter.

In order to smooth out the bursts and facilitate a comparison between 
the SG and LP models, we calculate the mean 
mass accretion rates $\langle \dot{M} \rangle$ by time-averaging the instantaneous rates over 
$2\times 10^4$~yr. Figure~\ref{fig5} shows $\langle \dot{M} \rangle$ versus time
for the SG model (solid line), $\alpha_{\rm LP}=10^{-4}$ LP model (dashed line),
$\alpha_{\rm LP}=10^{-2}$ LP model (dash-dotted line), and $\alpha_{\rm LP}=0.25$ LP model (dotted line).
The horizontal axis shows time elapsed since the formation of the central star.
The burst phenomenon is illustrated in the insert to Fig.~\ref{fig5}, which 
shows the instantaneous mass accretion rates in the SG model as a function of time.
It is seen that the time-averaged mass accretion rates are identical before the disk 
formation ($t\le 0.14$~Myr) but become distinct soon afterward.
Even after time-averaging, some substantial variations in the mass accretion rate 
of the SG model are still visible. 
It is evident that the $\alpha_{\rm LP}=10^{-4}$ LP model greatly underestimates 
$\langle \dot{M} \rangle$ in the early $0.4$~Myr, while considerably overestimating it in
the subsequent evolution. The high-viscosity $\alpha_{\rm LP}=0.25$ LP model 
also seem to produce larger accretion rates than those of the exact solution, 
especially in the late phase. The $\alpha_{\rm LP}=10^{-2}$ LP model, again, demonstrates
better agreement with the exact solution, though  somewhat underestimating $\langle \dot{M}\rangle$
after 1~Myr.

\subsection{The KMK model}
In this section we investigate the efficiency of the \possessivecite{Kratter} 
$\alpha$-parameterization in simulating the effect of GI in circumstellar disks.
The KMK model described by equations~(\ref{KMK})-(\ref{Kratter2})  is more useful and flexible 
than the LP model because the former has no free parameters\footnote{In fact, the value of the critical
Toomre parameter $Q_{\rm c}$ is set to 1.3 in the Kratter et al.'s approach.} and it 
includes a mild dependence on the disk-to-total mass ratio $\mu$.
This ratio is expected to depend on the initial rate of rotation of a molecular cloud core
and may considerably vary along the sequence of stellar evolution phases, and therefore
could be an important ingredient for successful modeling of the effect of GI 
in circumstellar disks. For a more detailed discussion we refer the reader to 
\citeasnoun{Kratter}.

\begin{figure}
  \resizebox{\hsize}{!}{\includegraphics{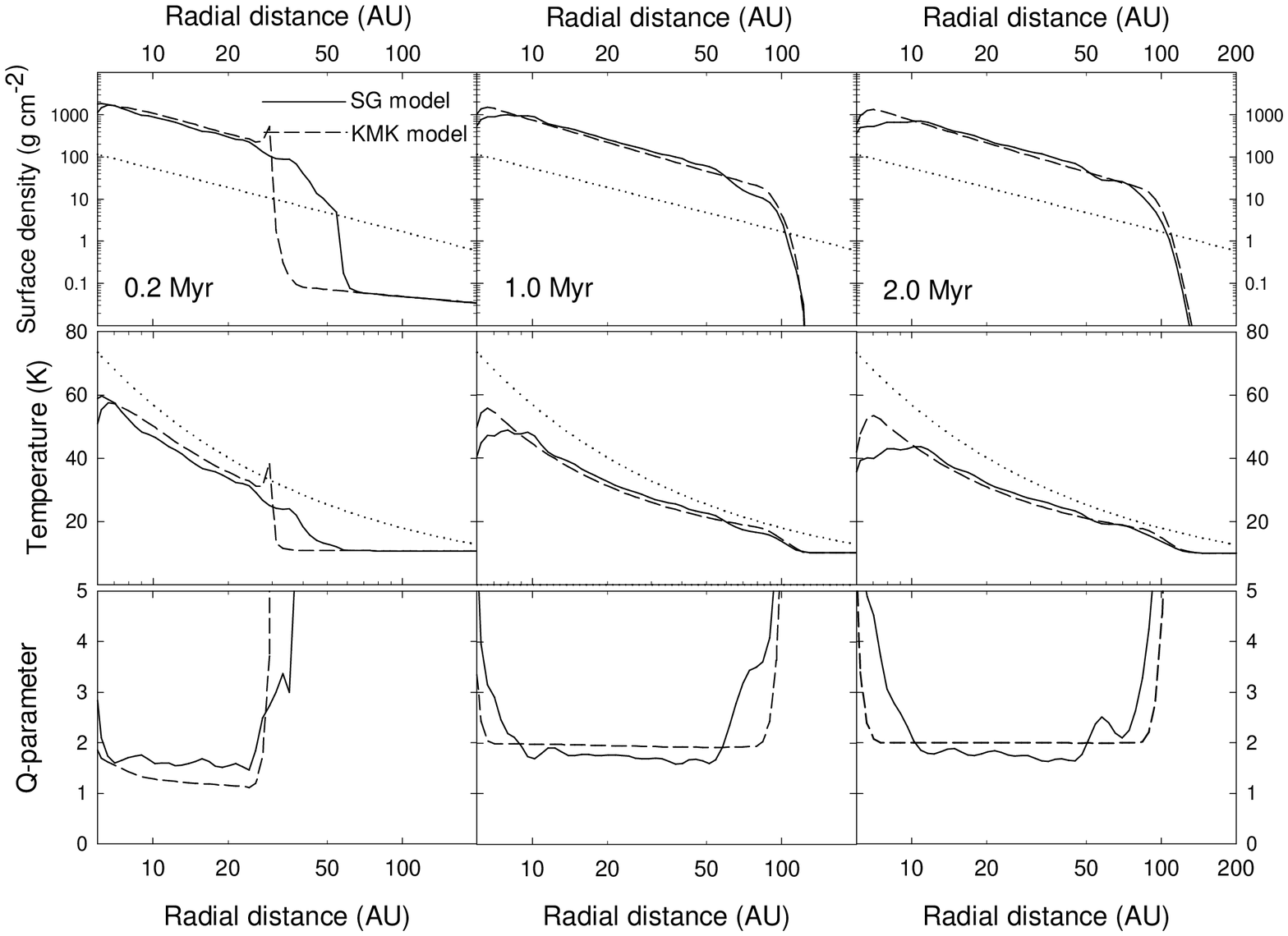}}
      \caption{Gas surface density (top row), gas temperature (middle row), and $Q$-parameter (bottom
      row) as a function of radius in the KMK model (dashed lines) and 
      SG model (solid lines). Three
      distinct evolutionary times since the formation of the central protostar are indicated in each
      column. The dotted lines show the MMSN gas surface density profile (top row) and 
      a typical gas temperature profile (middle row), as inferred fir circumstellar disks by AW05}
         \label{fig6}
\end{figure}

Figure~\ref{fig6} shows radial profiles of the gas surface density (top), temperature (middle),
and Toomre parameter (bottom) obtained in the KMK model 
(dashed lines) and in the SG model (solid lines). The horizontal axis, as usual, shows time 
elapsed since the formation of the central star. Three characteristic stellar ages,
as indicated in each row, are chosen for the presentation.
It is evident that the KMK model shows a satisfactory performance, particularly
in the late evolution stage. In the early evolution, the predicted disk surface densities and
temperatures are slightly larger than those of the SG model, but the difference is 
not significant. Comparing Figs~\ref{fig2} and \ref{fig6} one can see that the KMK model 
reproduces the exact solution to the same extent and accuracy as the best 
$\alpha_{\rm LP}=10^{-2}$ LP model.
However, the supremacy of the KMK model is obvious -- it has no free parameters to adjust.
We believe that the KMK model owes its success to the use of a two-component 
$\alpha$-parameter, $\alpha_{\rm GI}$.

The right column in Fig.~\ref{fig4} shows the stellar mass (top), disk mass (middle), and disk-to-star
mass ratio $\xi$ obtained using the KMK model (dashed lines) and SG model (solid lines). 
This figure demonstrates that the KMK model yields the disk and 
stellar masses that are in good agreement with those of the SG model. We conclude that both the LP and KMK models can, in principle, reproduce the
radial structure of self-gravitating disks, time-averaged accretion rates, and masses
in star/disk systems formed from slowly rotating  cloud cores with $\beta \la 1.0\times 10^{-3}$. 
Such systems are characterized by disk-to-star mass ratios not exceeding $\xi=0.2$. The 
$\alpha$-parameterization in such systems appears to be justified.
The agreement with the exact solution 
is somewhat modest in the early evolution but improves considerably in the late evolution
as the strength of gravitational instability declines.
In the case of LP model, the $\alpha$-parameter has to be set to $\alpha_{\rm LP}\approx 0.01$.

\section{Cloud cores with intermediate and high rates of rotation.}
In this section we consider cloud cores that are characterized by 
the ratios of rotational to gravitational energy $\beta_2=2.3\times 10^{-3}$
and $\beta_3=8.2\times 10^{-3}$, with most attention being concentrated on the latter case.
Since the KMK model has proven superior in comparison to the LP model, we focus
on the former one, summarizing the main results for the LP model where necessary.
Cloud cores with high rates of rotation are expected to yield disks with
large disk-to-star mass ratios. Our motivation is then to determine the extent to which the
KMK model can reproduce the exact solution in systems with disk-to-star mass ratios 
considerably larger than that of the previous section, $\xi\sim 0.17$. 

\begin{figure}
  \resizebox{\hsize}{!}{\includegraphics{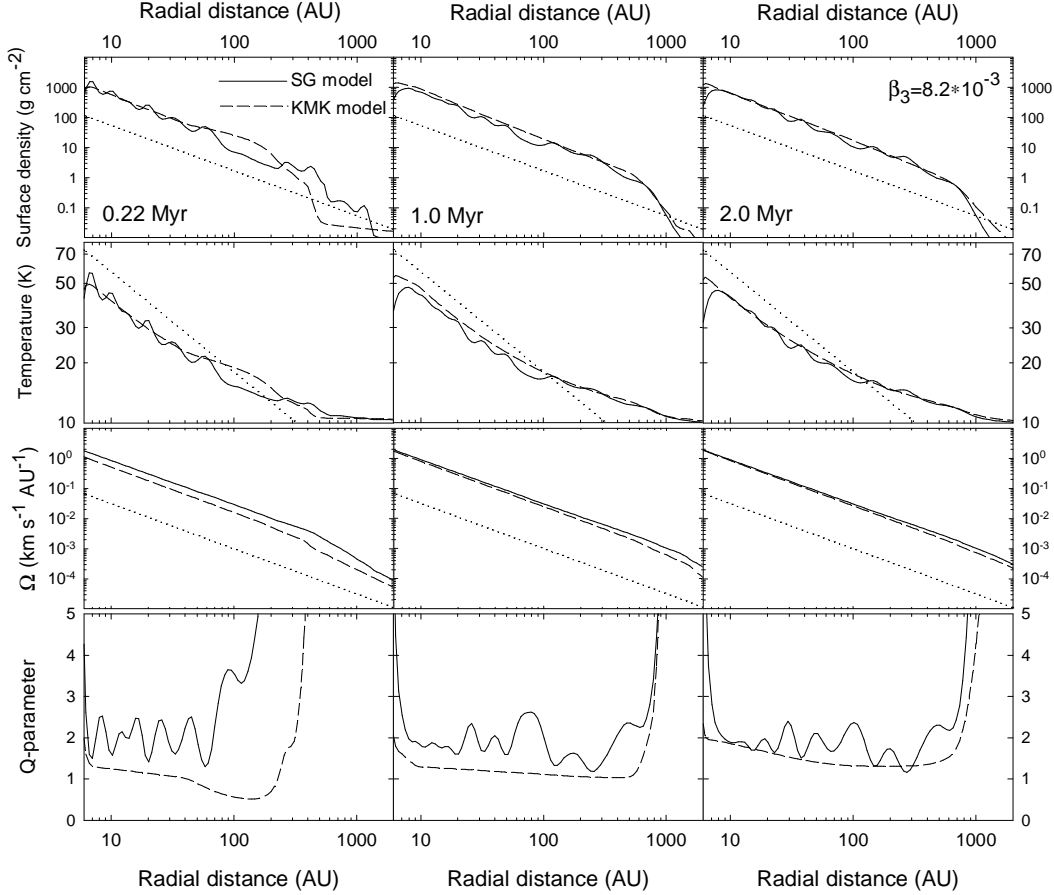}}
      \caption{From top to bottom: gas surface density, gas temperature, angular velocity, and 
      $Q$-parameter as a function of radius in the KMK model (dashed lines) and 
      SG model (solid lines) for the $\beta_3=8.2 \times 10^{-3}$ case. Three
      distinct evolutionary times since the formation of the central protostar are indicated 
      in each column.  The dotted lines show the MMSN gas surface density profile (first row), 
      typical gas temperature profile (second row) as inferred in circumstellar disks by AW05,
      and a Kepler rotation law (third row).}
         \label{fig7}
\end{figure}

Figure~\ref{fig7} compares the disk radial structure obtained in the SG~model
(solid lines) and KMK~model (dashed lines) for the $\beta_3$ case. The layout of the figure
is exactly the same as in Fig.~\ref{fig1},  but we have also plotted 
the radial distribution of the angular velocity $\Omega$ in the third row from top.
The dotted line in this row shows a Kepler rotation law, $\Omega \propto r^{-1.5}$.
 
Both the approximate and exact solutions are characterized by similar slopes of 
the gas surface density $\Sigma \propto r^{-1.5}$,
gas temperature $T \propto r^{-0.4}$, and angular velocity $\Omega \propto r^{-1.5}$.  
One can see that the KMK model reproduces the radial gas surface 
densities and temperatures but yields too low angular velocities and the Toomre parameter,
especially in the early evolution. 
This implies that the KMK model underestimates the mass of the central star.

\begin{figure}
  \resizebox{\hsize}{!}{\includegraphics{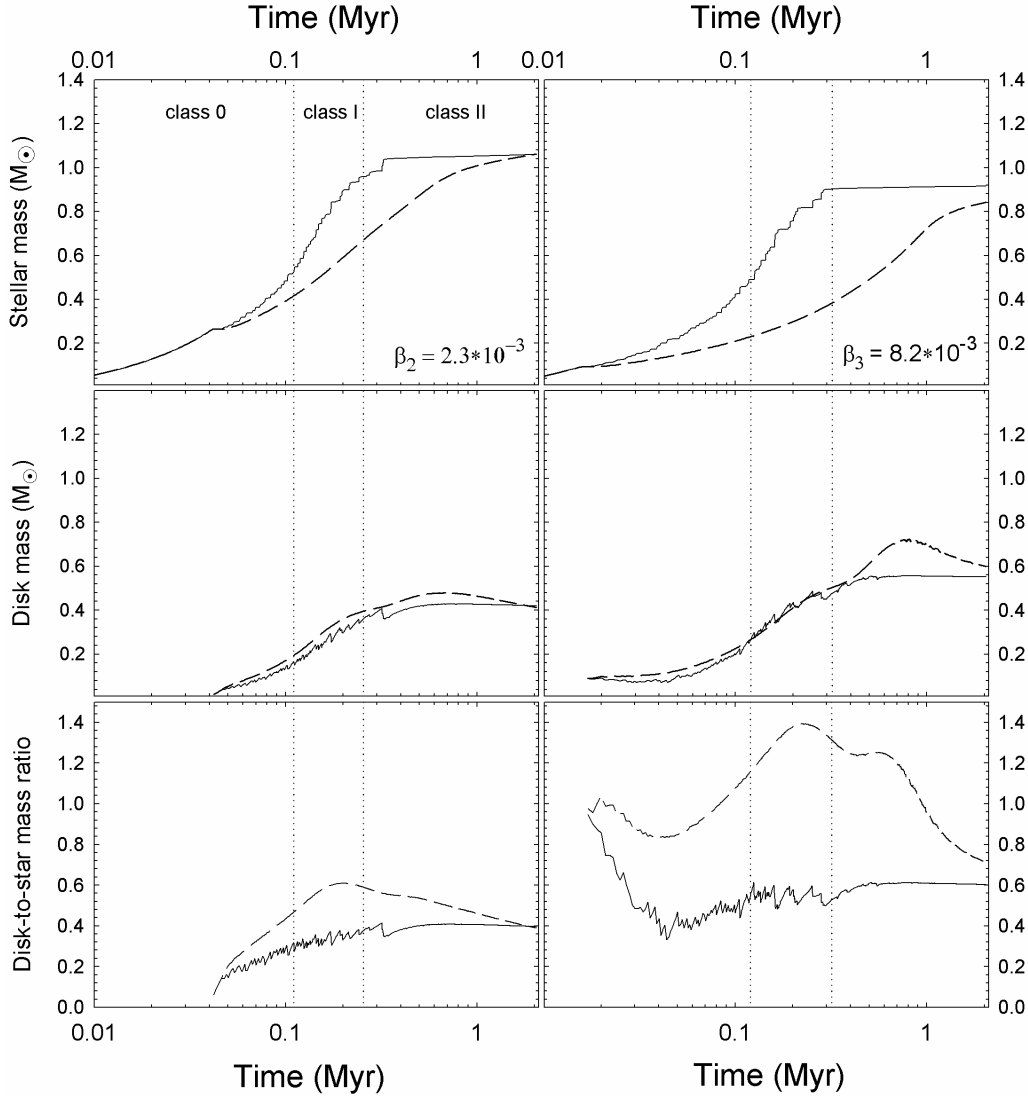}}
      \caption{Time evolution of stellar masses (top row), disk masses (middle row),
      and disk-to-star mass ratios (bottom row) in the KMK model (dashed lines) and SG~model
      (solid lines). The left and right columns correspond to the $\beta_2=2.3\times 10^{-3}$ 
      and $\beta_3=8.2\times 10^{-3}$ cases, respectively. The horizontal axis shows time since the
      formation of the central star. }
         \label{fig8}
\end{figure}

The failure of the KMK model to accurately reproduce stellar masses 
is illustrated in Fig.~\ref{fig8}, 
which show the stellar masses (top row), disk masses (middle row), and
disk-to-star mass ratios (bottom row) for the SG~model (solid lines)
and KMK model (dashed lines). In particular, the left column corresponds to the
$\beta_2$ case, while the right column presents the $\beta_3$ case.
The horizontal axis shows time elapsed since the formation of the central star.
To compare masses along the sequence of stellar evolution phases,
we need an evolutionary indicator to distinguish between Class 0, Class I, 
and Class II phases. 
We use a classification of \citeasnoun{Andre}, who
suggest that the transition between Class 0 and Class I objects occurs when
about $50\%$ of the initial cloud core is accreted onto the protostar-disk system.
The Class II phase is consequently defined by the time when the
infalling envelope clears and its total mass drops below $10\%$ of the initial cloud 
core mass $M_{\rm cl}$.
The vertical dotted lines mark the onset of Class I (left line) and Class~II (right line) phases.
A general trend of the KMK~model to {\it underestimate} the stellar masses
and to {\it overestimate} the disk-to-star mass ratios is clearly seen.

To quantify this mismatch between the models, we calculate time-averaged stellar
masses $\langle M_\ast \rangle$, disk masses $\langle M_{\rm d} \rangle$,
and disk-to-star mass ratios $\langle \xi \rangle$ in each major stellar evolution phase.
The resulted values are listed in Table~\ref{table1}.
It is evident that the mismatch between the SG model and KMK model is particularly strong in 
the Class 0 and Class I phases. For instance, the $\beta_3$ KMK model yields $\langle \xi \rangle=1.31$
in the Class I phase, which is almost a factor of 3 larger than the corresponding value for the SG model.
The disagreement in the Class II phase is in general less intense than in the Class 0 and Class I phases.
Somewhat surprisingly, disk masses in the KMK model differ 
insignificantly from those of the SG model. 
The $\alpha_{\rm LP}=10^{-2}$ model shows a very similar behaviour.

\begin{table}
\caption{SG model versus KMK model for $\beta_2=2.3\times 10^{-3}$  and $\beta_3=8.2\times10^{-3}$
\label{table1}}
\begin{tabular}{cc}
\begin{tabular}{cccc}
\hline\hline
phase & $\langle M_\ast \rangle_{\beta_2}$ & $\langle M_{\rm d} \rangle_{\beta_2}$ & 
$\langle \xi \rangle_{\beta_2}$ \\
\hline
 Class 0  & 0.39/0.33 & 0.09/0.12  & 0.23/0.33  \\
 Class I  & 0.81/0.56 & 0.28/0.33  & 0.34/0.58  \\
 Class II & 1.05/0.98 & 0.42/0.44  & 0.4/0.46   \\
 \hline
\end{tabular} 
\begin{tabular}{ccc}
\hline\hline
$\langle M_\ast \rangle_{\beta_3}$ & $\langle M_{\rm d} \rangle_{\beta_3}$ & 
$\langle \xi \rangle_{\beta_3}$ \\
\hline
 0.29/0.17 & 0.14/0.17  & 0.49/0.97  \\
 0.77/0.31 & 0.42/0.42  & 0.54/1.33  \\
 0.91/0.71 & 0.55/0.64  & 0.6/0.94   \\
 \hline
\end{tabular} 
\end{tabular}
\\All mean masses are in $M_\odot$. The slash differentiates between the SG model (left) and
KMK model (right).
\end{table} 

\begin{figure}
\includegraphics[width=12 cm]{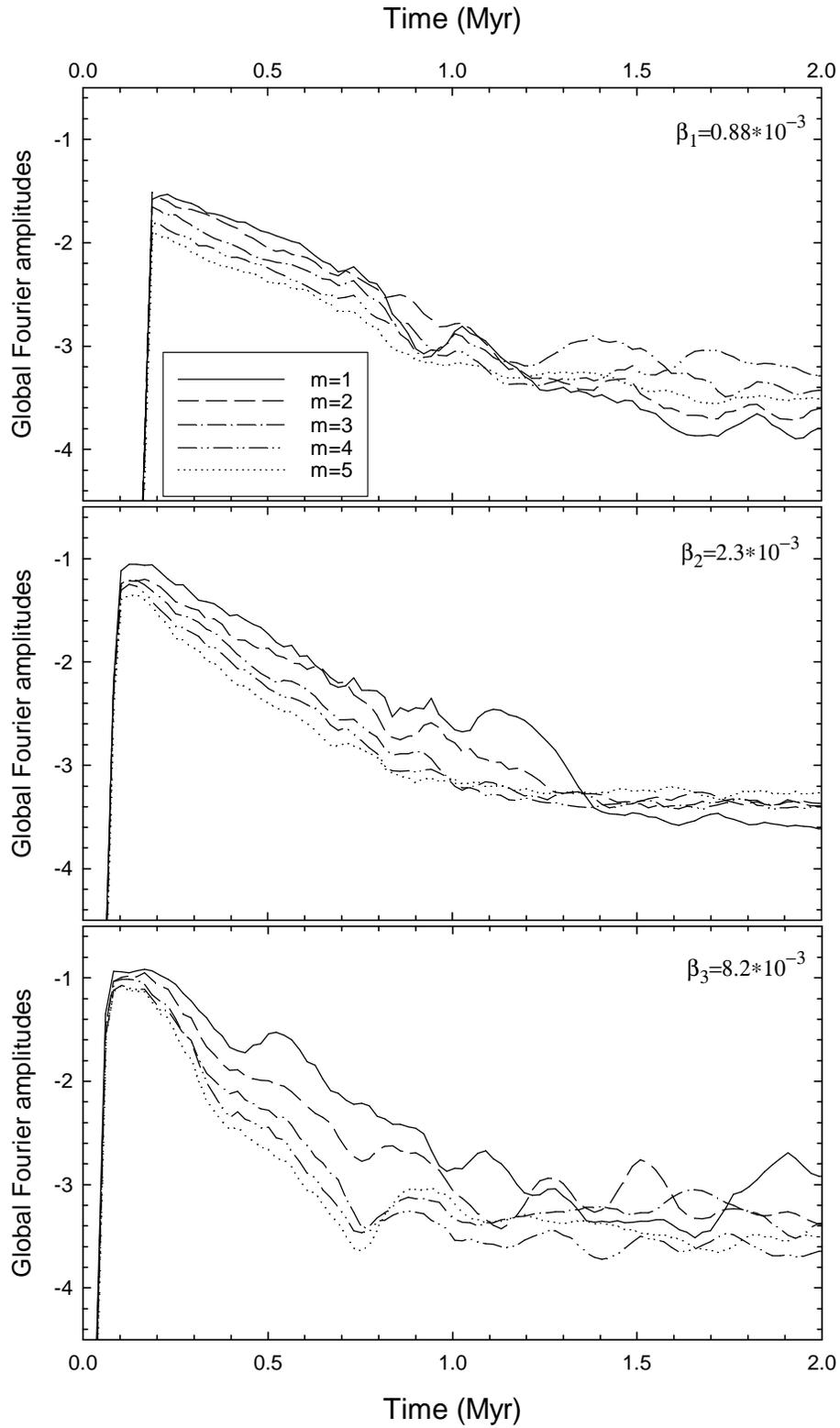}
      \caption{Global Fourier amplitudes (averaged over $2\times10^{4}$~yr)
      in the $\beta_1=0.88\times10^{-3}$ SG model (top), $\beta_2=2.3\times 10^{-3}$ SG model 
      (middle), and $\beta_3=8.2\times 10^{-3}$ SG model
      (bottom). The horizontal axis shows time elapsed since the formation of the central
      star. Each line type corresponds to a mode of specific order, as indicated in the legend. }
         \label{fig9}
\end{figure}

Our numerical simulations demonstrate that the extent to which the 
KMK model departs from the SG model is direct proportional to the value of $\beta$, 
and hence to the disk-to-star mass ratio. This is not surprising. Disks in systems with 
greater $\xi$ are more gravitationally unstable, and the viscous approach is expected to fail 
in massive disks, which are dominated by global spiral modes of low order 
\citeaffixed{Lodato05,Krumholz}{see e.g.}.
We illustrate this phenomenon by calculating the global Fourier amplitudes (GFA) defined
as 
\begin{equation}
C_{\rm m} (t) = {1 \over M_{\rm d}} \left| \int_0^{2 \pi} \int_{r_{\rm in}}^{r_{\rm disc}} 
\Sigma(r,\phi,t) \, e^{im\phi} r \, dr\,  d\phi \right|,
\end{equation}
where $M_{\rm d}$ is the disc mass and $r_{\rm disc}$ is the disc's physical outer radius.
The instantaneous GFA show considerable fluctuations and we have to time-average 
them over $2\times10^4$~yr in order to produce a smooth output.
The time evolution of the time-averaged GFA (log units) is shown in Fig.~\ref{fig9} 
for the $\beta_1=0.88\times 10^{-3}$ SG model (top), $\beta_2=2.3\times 10^{-3}$ SG model (middle),
and $\beta_3=8.2\times 10^{-3}$ SG model (bottom). Each color type corresponds to a mode 
of specific order, as indicated in the legend

The time behaviour of the GFA is indicative of two qualitatively different stages in the disk 
evolution. In the early stage ($t\la 1.0$~Myr),
a clear segregation between the modes is evident -- the lower order mode dominates its 
immediate higher order neighbour in all models. In particular, the $m=1$ mode is almost always 
the strongest one\footnote{In the present paper, we have ignored a possible wobbling of the central
star, which may increase the strength of the odd modes, especially that of the $m=1$ mode 
\cite{Adams89,Shu90}. Numerical hydrodynamics simulations with the indirect potential
in the momentum equation~(\ref{visc2}) are needed to accurately assess the strength of this effect.}. The modes also show a clear tendency to decrease in magnitude with time.
In the late stage, however, this clear picture breaks into a kaleidoscope of low-magnitude 
modes competing for dominance with each other. 
These mode fluctuations are not a numerical noise but are rather 
caused by ongoing low-amplitude
non-axisymmetric density perturbations sustained by swing amplification at the disk's sharp outer edge.
As was shown by \citeasnoun{VB3}, self-gravity of the disk is essential for these 
density perturbations to persist into the late disk evolution. The density perturbations quickly 
disappear if self-gravity is switched off.

The visual analysis of GFA in the early stage reveals that models with greater $\beta$ are 
characterized by spiral modes of greater magnitude. For instance, the magnitude of the 
$m=1$ mode at $t=0.5$~Myr in the $\beta_1$ model is $C_1=-1.9$~dex, while
$\beta_2$ and $\beta_3$ models have $C_1=-1.7$~dex and $C_1=-1.5$~dex, respectively. 
What is more important is that the relative strength of the low-order modes ($m\le2$) is greater 
in models with greater $\beta$ (and greater $\xi$). These facts, when taken altogether, 
account for the failure of the viscous approach in systems with $\xi\ga0.2$.
The preponderance of strong low-order modes, which are global in nature, largely 
invalidates the local-in-nature viscous approach.
When we turn to the late stage at $t\ga 1$~Myr, we see that all modes saturate at a considerably lower
value of order $C=-3.5$~dex. Moreover, characteristic fluctuations tend to produce more 
cancellation in the net gravitational torque on large scales, thus making the GI-induced transport 
localwise. This explains why the KMK model performs better in the late stage.

\begin{figure}
\resizebox{\hsize}{!}{\includegraphics{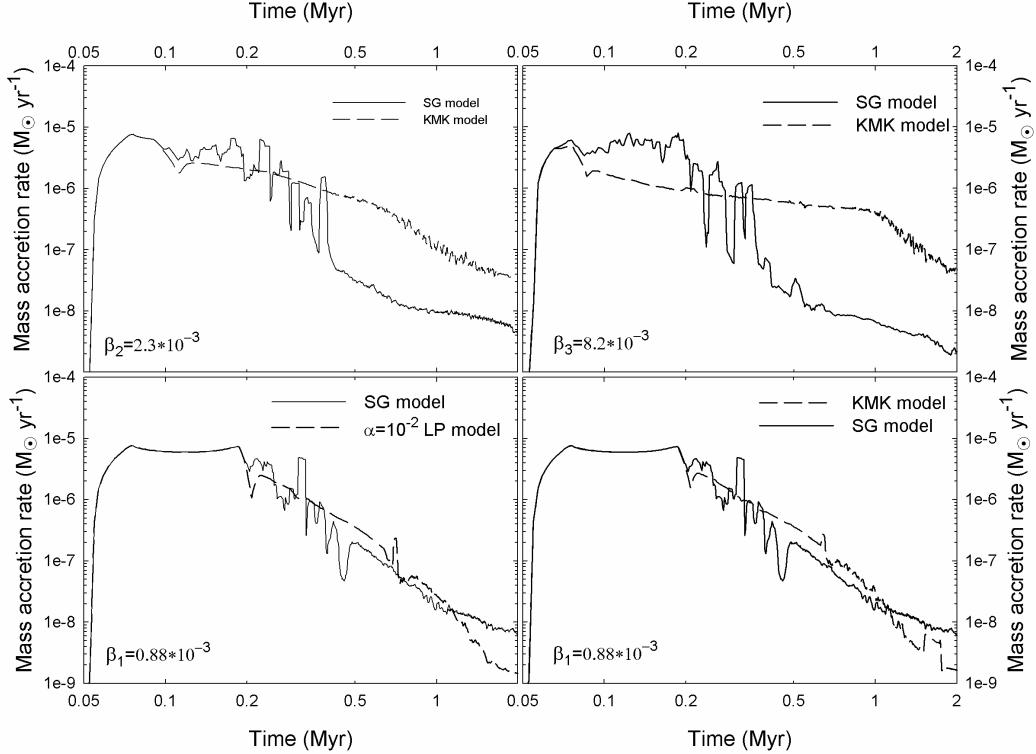}}
      \caption{Mass accretion rates $\langle \dot{M} \rangle$  
      (averaged over $2\times 10^4$~yr) versus time.
      The top row shows $\langle \dot{M} \rangle$ for $\beta_2=2.3\times 10^{-3}$ (top-left panel)
      and $\beta_3=8.2\times 10^{-3}$ (top-right panel). In particular, the solid and dashed lines correspond
      to the SG and KMK models, respectively. The bottom row presents
      $\langle \dot{M} \rangle$ for the $\beta_1=0.88\times 10^{-3}$ case. 
      More specifically, the dashed lines in the left and right panels show the data 
      for the $\alpha_{\rm LP}=10^{-2}$ LP model and KMK model, respectively, while the solid lines
      in both low panels present the data for the SG model.   }
         \label{fig10}
\end{figure}

Finally, we present in Figure~\ref{fig10} the time-averaged  
mass accretion rates $\langle \dot{M} \rangle$ as a function of time for the 
$\beta_2=2.3\times 10^{-3}$ case (top-left panel) 
and $\beta_3=8.2\times 10^{-3}$ case (top-right panel). In particular, the solid and 
dashed lines correspond to the SG model and KMK model, respectively.
The time-averaged values are obtained from the instantaneous mass accretion rates by
applying a running average method over $2\times 10^{4}$~yr.
For comparison, the bottom row presents $\langle \dot{M} \rangle$ versus time for the 
$\beta_1=0.88\times 10^{-3}$ case. More specifically, the dashed line in the bottom-left panel corresponds
to the best $\alpha_{\rm LP}=10^{-2}$ LP model, while the same line type in the bottom-right
panel presents the KMK model. In both bottom panels, the solid line shows the data for the SG model.
The horizontal axis shows time elapsed since the beginning of numerical simulations. 

The central star
forms in all models at around 0.05 Myr, when $\langle \dot{M} \rangle$ increases sharply to a maximum
value of $\sim 10^{-5}~M_\odot$~yr$^{-1}$. A period of near constant accretion then ensues, when
the matter is accreted directly from the infalling envelope onto the star. This stage
may be very short (or even evanescent) in systems with high rates of rotation (top panels) due to 
almost instantaneous onset of a disk formation stage. In this stage,
the matter is first accreted onto the disk and through the disk onto the star, and
the mass accretion rate is highly variable. It is evident
that the $\beta_2$ and $\beta_3$ KMK models (top row) underestimate $\langle \dot{M}\rangle$ in 
the early disk evolution, while considerably overestimating $\langle \dot{M}\rangle$ in the late evolution.
This explains why viscous models with large $\xi$ tend to greatly underestimate
stellar masses in the Class 0 and Class I phases.
It appears that viscous torques in massive disks fail to keep up with gravitational torques in 
the early, strongly gravitationally 
unstable phase but drive too high rates of accretion in the late, gravitationally quiescent phase.
On the other hand, the bottom row in Fig.~\ref{fig10} indicates that 
viscous models with low rates of rotation and small disk-to-star mass ratios 
($\beta\la 1\times 10^{-3}$ and $\xi\la 0.2$) show a better fit to the SG model, 
departing from the exact solution only by a factor of several.

\section{Higher disk temperature}
\label{hotdisk}
Observations and numerical simulations suggest that circumstellar disks are characterized 
by a variety of physical conditions,
including vastly different disk sizes, disk-to-star mass ratios, and temperature profiles 
\citeaffixed{Andrews05,Vor09}{see e.g.}.
We have considered the effect of different disk-to-star mass ratios in the previous 
sections. However, it is also important
to consider disks with different temperatures, since this is one of the factors 
that control the disk propensity to gravitational
instability and fragmentation.  In our polytropic approach, we can increase the disk temperature 
by raising the ratio of specific heats $\gamma$. In a real disk, the rise in $\gamma$ does not
necessarily result in higher temperatures due to a strong dependence on cooling and heating terms.
These effect are not considered in the present study.  

\begin{figure}
  \resizebox{\hsize}{!}{\includegraphics{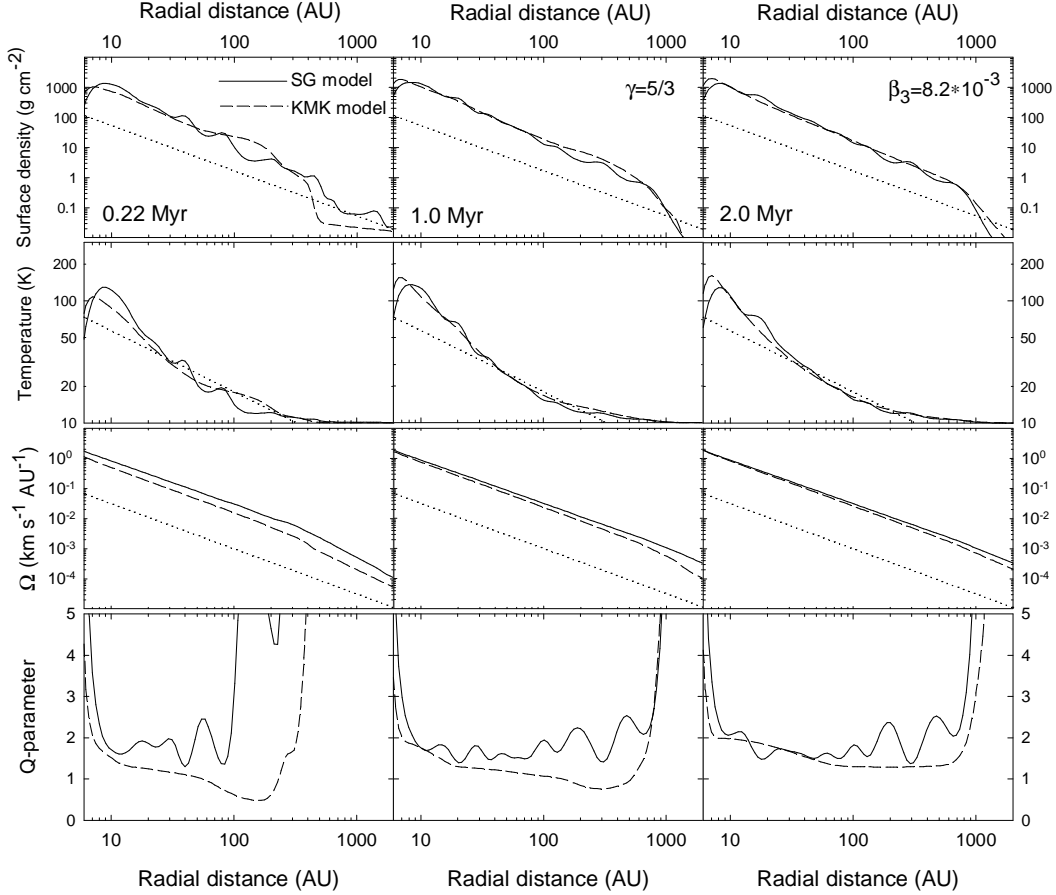}}
      \caption{The same as Figure~\ref{fig7} only for the $\gamma=5/3$ disk.}
         \label{fig11}
\end{figure}

Figure~\ref{fig11} presents the disk radial structure for the 
$\gamma=5/3$ and $\beta_3=8.2\times 10^{-3}$ case.
The dashed and solid lines show the data obtained in the KMK model and SG~model, respectively.
The layout of the figure is identical to that of Fig.~\ref{fig7}. The comparison of Fig.~\ref{fig7}
and Fig.~\ref{fig11} reveals that the $\gamma=5/3$ disk is characterized by roughly a factor of 2 greater
gas temperature than that of the $\gamma=1.4$ disk. In fact, the $\gamma=5/3$ disk has temperatures
that systematically exceed typical temperatures inferred by AW05 for a sample of circumstellar 
disks (dotted line 
in the second row). Nevertheless, the performance of the KMK model is largely unaffected
by this change in the disk temperature. As in the $\gamma=1.4$ case, the KMK model yields too low angular
velocities in the early evolution (third row), thus underestimating stellar masses. 
In general, the temporal behaviour of disk masses, 
stellar masses, and disk-to-star mass ratios is very similar to that shown in Fig.~\ref{fig8} for the
$\gamma=1.4$ disk. This is not unexpected. Disk and stellar
masses are largely determined by the mass accretion rate onto the star.
As \citeasnoun{VB4} have demonstrated, a factor of 2 increase in the disk temperature
makes little effect on the time-averaged mass accretion rates, though it may 
alter the temporal behaviour of the instantaneous mass accretion rates.

\section{Disk fragmentation}
\label{fragment}
When the local Toomre paramter $Q_{\rm T}$ drops below some critical value $Q_{\rm cr,f}$, 
which is often equal 
or close to 1.0, fragmentation in a circumstellar disk ensues. To account for
this qualitatively different regime of disk evolution, \citeasnoun{Kratter} have 
defined a critical surface density 
\begin{equation}
\Sigma_{\rm cr,f}= {c_{\rm s} \Omega \over \pi G Q_{\rm cr,f} }
\end{equation}
and assumed that fragmentation depletes the disk surface density when 
$Q_{\rm T}<Q_{\rm cr,f}$ at a rate
\begin{equation}
\dot{\Sigma}_{\rm f}=-(\Sigma - \Sigma_{\rm cr,f}) \Omega.
\end{equation}
According to \citeasnoun{Kratter}, this rate is fast enough to ensure that 
$Q_{\rm T}$ never dips appreciably below $Q_{\rm cr,f}$.
The actual dynamics of the fragments is not followed, instead they are allowed 
to accrete on to the star at a rate $\dot{M}_{\ast,f}=0.05 \, M_{\rm f} \, \Omega$.

This approach, feasible in model simulations akin to those presented
by \citeasnoun{Kratter},  is difficult to implement in numerical 
hydrodynamic simulations. Taking away matter and instantaneously transporting
it over a large distance in the numerical grid may lead to numerical 
instabilities of unknown consequences. One possible way around this problem is
to actually create fragments and follow their dynamics. However, when disk self-gravity
is absent, this approach is also misleading because the dynamics of such fragments
is largely governed by the gravitational interaction with the disk. 

On the other hand, our viscous models demonstrate that the fragmentation regime
is sometimes achieved in the early disk evolution (see e.g. Figure~\ref{fig7}) 
and some changes to the standard effective-viscosity approach are necessary.
In the present work, we modify the $\alpha$-parameter
so as to mimic a possible increase in the efficiency of mass transport 
due to fragmentation. In particular, in the \possessivecite{Kratter}
formulation of $a_{\rm GI}$ we have made the following modification to
the short component
\begin{equation} 
\alpha_{\rm short,f} = \alpha_{\rm short}  \left( {Q_{\rm cr,f} \over Q_{\rm T}}\right)^n,  
\end{equation}
if the fragmentation regime with $Q_{\rm T}\le Q_{\rm cr,f}$ is set in the disk. 
In the usual regime of $Q_{\rm T}>Q_{\rm cr,f}$, the short component is not modified, 
i.e. $\alpha_{\rm short,f} = \alpha_{\rm short}$.  We set $Q_{\rm cr,f}=1.0$ and 
$n=10$, thus introducing a strong non-linearity in the expression for 
the $\alpha$-parameter. We note that in the definition of $\alpha_{\rm short}$ 
given by Equation~(\ref{Kratter1}), $Q_{\rm T}$ is not allowed to drop below unity by
setting $Q_{\rm T}=\max(Q_{\rm T},1.0)$. 
In practice, this means that $\left( {Q_{\rm cr,f}/Q_{\rm T}}\right)^n$ is the only 
term that is sensitive to fragmentation.


\begin{figure}
  \resizebox{\hsize}{!}{\includegraphics{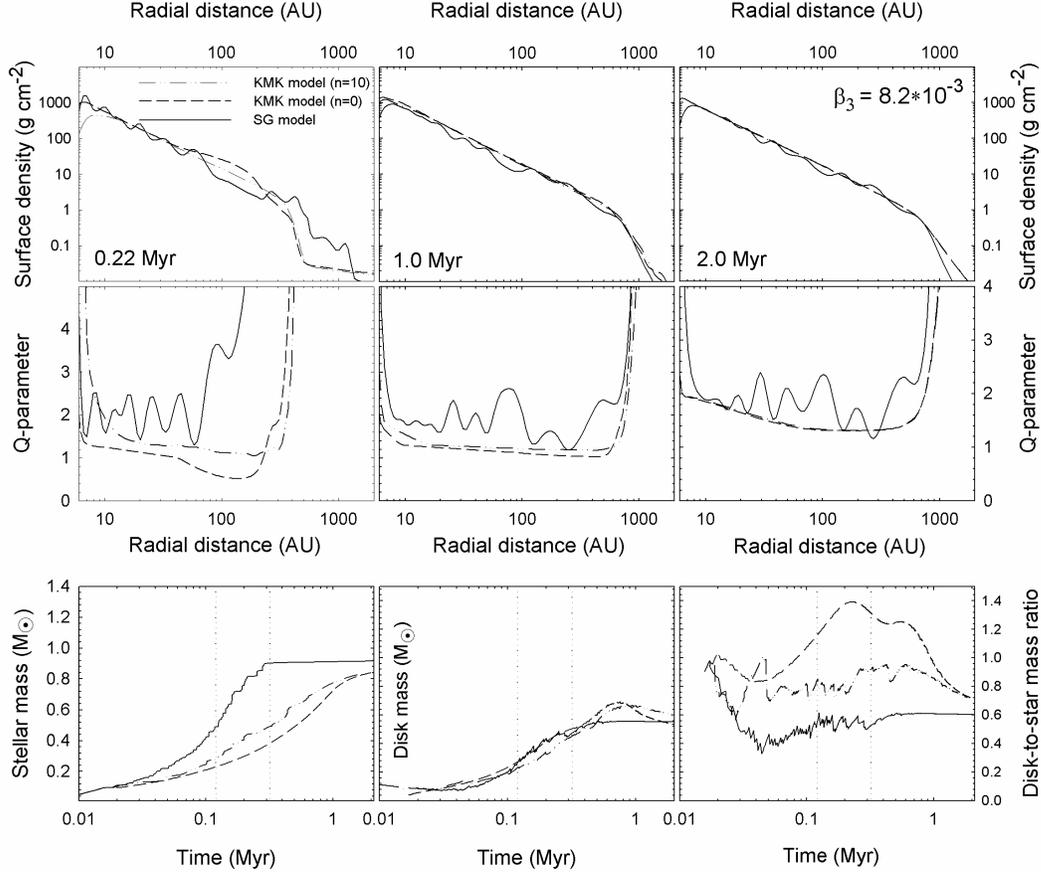}}
      \caption{Radial gas surface density profiles (top row) and Toomre Q-parameters
      (middle row) in the SG model (solid lines), unmodified KMK model (dashed lines),
      and modified-for-fragmentation KMK model (dash-dot-dotted lines). The bottom row
      presents the time evolution of the stellar mass (left panel), disk mass (middle panel),
      and disk-to-star mass ratio (right panel) in all three models. }
         \label{fig12}
\end{figure}

\begin{figure}
  \resizebox{\hsize}{!}{\includegraphics{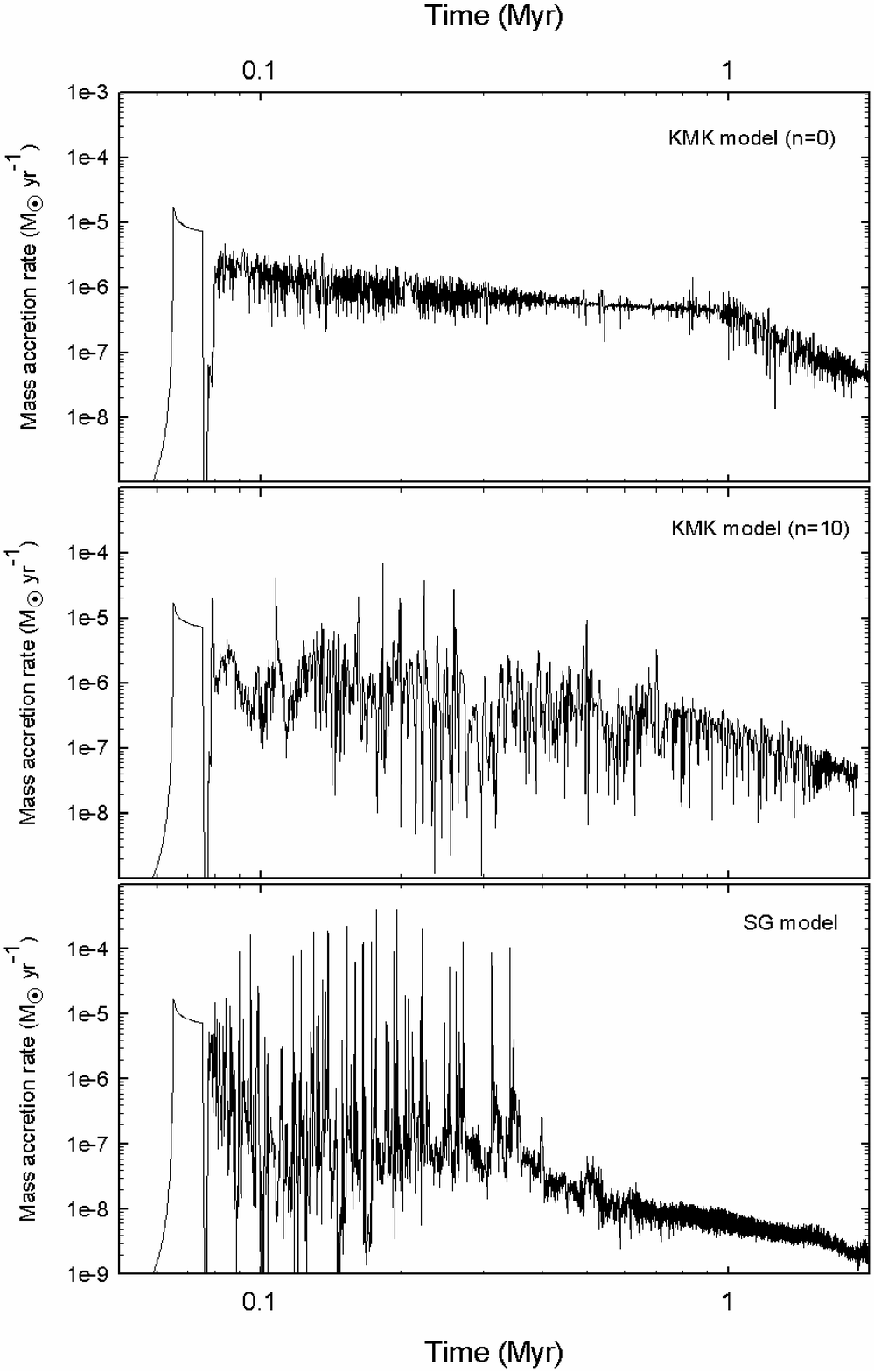}}
      \caption{Time evolution of instantaneous mass accretion rates in the unmodified KMK model (top
      panel), modified-for-fragmentation KMK model (middle panel), and SG model (bottom panel). }
         \label{fig13}
\end{figure}

Figure~\ref{fig12} presents a comparison of the KMK models with and without 
fragmentation with the SG model for the case of $\beta=8.2\times 10^{-3}$ 
and $\gamma=1.4$. The top and middle rows show the radial profile snapshots of the 
gas surface density and Q-parameter
at three different times after the formation of the central star.
The bottom row shows the time evolution
of the stellar mass (left panel), disk mass (middle panel), and disk-to-star mass 
ratio (right panel).
As usual, the solid and dashed lines represent the SG~model and unmodified KMK model,
respectively. The dash-dot-dotted line shows the KMK model modified for fragmentation.
The vertical dotted lines mark the onset of the Class I and Class II phases in the 
SG model.

There are several interesting features in Figure~\ref{fig12} that deserve attention.
When fragmentation is taken into account in the KMK model, 
the Q-parameter never dips appreciably below the critical value for fragmentation 
$Q_{\rm cr,f}=1.0$, contrary to what was sometimes seen in the unmodified KMK model.
This illustrates a self-regulating nature of the modification we have made to the standard
KMK model. Furthermore, the modified KMK model reproduces better the exact solution,
though the mismatch is still substantial. For instance, the disk-to-star mass ratio
in the modified KMK model is much closer to that of the SG model. 
In particular, the disk in the modified KMK model is almost always less massive than the star, 
which is in agreement with the expectations of the exact SG model \cite{Vor09}.

Perhaps, the most important feature of the modified KMK model
is a non-monotonic, step-like increase in the mass of the central star with time, 
akin to that of the SG model. The disk-to-star mass ratios also demonstrates 
temporal variations similar to those of the SG model. We emphasize that the unmodified
KMK model lacks such behaviour. These short-term variations are 
signatures of the mass accretion burst phenomenon. To illustrate this, we plot
the instantaneous mass accretion rates $\dot{M}$ versus time in Figure~\ref{fig13}  
for the unmodified KMK model (top), modified KMK model (middle), and SG model (bottom).
The mass accretion rate in the unmodified KMK model exhibits only small-amplitude 
flickering, there is no trace of the several-orders-of-magnitude variability that
is typical for the burst phenomenon.
Much to our own surprise, the modified KMK model was found to have short-term 
variations in $\dot{M}$ that are similar in amplitude to those of the SG model. 
Although the number of such bursts in the modified KMK model is still smaller
than in the SG model, the mere fact that, after a simple modification, 
the KMK model can reproduce the burst phenomenon and accretion variability
by several orders of magnitude is fascinating and encouraging.

\section{Conclusions}
We have performed numerical hydrodynamic simulations of the self-consistent
formation and long-term (2 Myr) evolution of circumstellar disks with the
purpose to determine the applicability of the viscous $\alpha$-parameterization
of gravitational instability in self-gravitating disks.
In total, we have considered three numerical models: 
the LP model that uses the \possessivecite{Lin90} $\alpha$-parameterization, 
the KMK model that adopts the \possessivecite{Kratter} $\alpha$-parameterization, 
and the SG model that employs 
no effective viscosity but solves for the gravitational potential directly (the exact solution). 
We then perform a detailed analysis of the resultant circumstellar radial disk structure, 
masses, and mass accretion rates in the three models 
for systems with different disk-to-star mass ratios $\xi$. We find the following. 
\begin{enumerate}
\item The agreement between the viscous $\alpha$-models and the SG model 
depends  on the value of $\xi$ and deteriorates
along the sequence of increasing disk-to-star mass ratios. In principle, the viscous
$\alpha$-models can provide an acceptable fit to the SG model for stellar systems
with $\xi\la0.2-0.3$, which is in agreement with $\xi\le0.25$ previously reported by
\citeasnoun{Lodato04} based on a different $\alpha$-parameterization. 

\item The success of the Lin \& Pringle's $\alpha$-parameterization in systems with 
$\xi\la0.2-0.3$ depends 
crucially on the proper choice of $\alpha_{\rm LP}$. The $\alpha_{\rm LP}=10^{-2}$ model 
yields an acceptable
fit to the exact solution but completely fails for $\alpha_{\rm LP}\la 10^{-4}$ 
and $\alpha_{\rm LP}\ga0.25$.
In particular, the $\alpha_{\rm LP}=10^{-4}$ model yields  a disk being too small
in size, having too large gas surface density and disk-to-star mass ratio.
On the other hand, the $\alpha_{\rm LP}=0.25$ model drives too large accretion rates, 
resulting in a disk being
heavily depleted in mass and having too small gas-to-star mass ratios already after 
1.0~Myr of evolution.

\item The performance of the KMK model is comparable to or even better than 
that of the best $\alpha_{\rm LP}=10^{-2}$ LP model. In addition, the former 
is superior because it has no explicit dependence on the $\alpha$-parameter,
and it includes some dependence on the disk-to-star mass ratio.

\item The viscous $\alpha$-models generally fail in stellar systems with 
$\xi \ga 0.3$. They yield too small stellar masses and too large  
disk-to-star mass ratios, especially in the 
early Class 0 and Class I evolution phases  (see Table~\ref{table1}). 
For instance, the KMK model may overestimate the disk-to-star mass ratio by a factor of 3
as compared to that of the SG model.
The same lack of agreement is also seen in the time-averaged mass accretion rates
$\langle \dot{M}\rangle$. In particular, 
the KMK model underestimates $\langle \dot{M}\rangle$ in the Class 0 and Class I phases,
while greatly overestimating $\langle \dot{M} \rangle$ in the Class II phase.

\item The failure of the KMK model (and LP model) in the case of large $\xi$ is related to 
the growing strength of low-order spiral modes in massive self-gravitating disks.

\item A simple modification to the $\alpha$-parameter that takes 
into account disk fragmentation can somewhat improve the performance of the KMK model
and even reproduce to some extent the mass accretion burst phenomenon \cite{VB1,VB2},
demonstrating the importance of the proper treatment for disk fragmentation. 
More numerical study is needed to explore this issue. 

\item Both viscous $\alpha$-models perform better in the late 
disk evolution (Class II phase) than in the early disk evolution (Class 0 and Class I phases),
irrespective of the value of $\xi$. This is due to a gradual decline in the magnitude
of the spiral modes with time, as well as due to growing mode-to-mode interaction.
The latter tends to produce more cancellation in the net gravitational torque on large scales,
thus making the effect of gravitational torques similar to that of local viscous torques.

\item A factor of 2 increase in the disk temperature does not noticeably affect the efficiency
of the viscous $\alpha$-models in simulating the effect of GI.
\end{enumerate}

\section*{Acknowledgements}
The author is thankful to the anonymous referee for an insightful report that
helped to improve the manuscript and to Prof. Shantanu Basu for stimulating discussions. The author
gratefully acknowledges present support from an ACEnet Fellowship. Numerical 
simulations were done on the Atlantic Computational  Excellence Network (ACEnet).

\end{document}